\documentclass[10pt,journal,compsoc]{IEEEtran}

\ifCLASSOPTIONcompsoc
\usepackage[nocompress]{cite}
\else
\usepackage{cite}
\fi
\usepackage{url}
\usepackage{xspace}
\usepackage{graphicx}
\usepackage{mathtools}
\usepackage{xcolor}
\usepackage{subcaption}
\usepackage{caption}
\usepackage{listings}
\usepackage{multirow}
\usepackage{booktabs}
\usepackage{threeparttable}
\usepackage{array}
\usepackage[most]{tcolorbox}
\usepackage[colorlinks = true,
            linkcolor = black,
            urlcolor  = black,
            citecolor = black,
            anchorcolor = black]{hyperref}
\newcolumntype{P}[1]{>{\centering\arraybackslash}p{#1}}
\definecolor{codegreen}{rgb}{0,0.6,0}
\definecolor{codegray}{rgb}{0.5,0.5,0.5}
\definecolor{codepurple}{rgb}{0.58,0,0.82}
\definecolor{backcolour}{rgb}{0.95,0.95,0.92}
 
\lstdefinestyle{mystyle}{
    commentstyle=\color{codegreen},
    keywordstyle=\color{purple},
    numberstyle=\tiny\color{codegray},
    stringstyle=\color{violet},
    basicstyle=\ttfamily\footnotesize,
    breakatwhitespace=false,         
    breaklines=true,                 
    captionpos=b,                    
    keepspaces=true,                 
    showspaces=false,                
    showstringspaces=false,
    showtabs=false,                  
    tabsize=2,
    frameround=fttt,
    xleftmargin=10pt,
    xrightmargin=10pt,
    language=Java
}
 
\usepackage{makecell}
 \usepackage{adjustbox}
\newcommand{\sql}[1]{\textcolor{black}{#1}}
\newcommand{\liu}[1]{\textcolor{black}{#1}}
\newcommand{\sq}[1]{\textcolor{black}{#1}}

\newcommand{\tool}{{{{ATOM}}}\xspace} 
\newcommand{\subtool}{{{\textit{AST2seq}}}\xspace}
\ifCLASSINFOpdf
\else
\fi
\usepackage{hyperref}
\usepackage{balance}
\begin{document}
\title{\tool: Commit Message Generation Based on Abstract Syntax Tree and Hybrid Ranking}

\author{Shangqing Liu, 
        Cuiyun Gao,
        Sen Chen,
        Lun Yiu Nie,
        and Yang Liu % <-this % stops a space
\IEEEcompsocitemizethanks{\IEEEcompsocthanksitem Shangqing Liu, Cuiyun Gao, Sen Chen, and Yang Liu are with School of Computer Science and Engineering, Nanyang Technological University, Singapore. \protect
E-mail: shangqin001@e.ntu.edu.sg, \{cuiyun.gao, chensen, yangliu\}@ntu.edu.sg
\IEEEcompsocthanksitem Lun Yiu Nie is with The Chinese University of Hong Kong, China. E-mail:lynie8@cse.cuhk.edu
\IEEEcompsocthanksitem Cuiyun Gao is the corresponding author.
}% <-this % stops an unwanted space
\thanks{Manuscript received December 7, 2019}}

\markboth{Journal of \LaTeX\ Class Files,~Vol.~14, No.~8, December~2019}%
{Liu \MakeLowercase{{et al.}}: \tool: Code Commit Message Generation Based on Hybrid Ranking}

\IEEEtitleabstractindextext{%
\begin{abstract}
Commit messages {record} code changes ({e.g.}, feature {modifications} and bug repairs) in natural language, and are useful for program comprehension. Due to {the} frequent updates of software and time cost, developers are generally unmotivated to write commit messages for code changes. Therefore, automating the message writing process is necessitated. 
Previous studies on commit message generation have been benefited from generation models or retrieval models, but the code structure of changed code, \sq{i.e., AST}, which can be important for capturing code semantics, has not been explicitly involved. Moreover, although generation models have the advantages of synthesizing commit messages for new code changes, they are not easy to bridge the semantic gap between code and natural languages which could be mitigated by retrieval models.
In this paper, we propose a novel commit message generation model, named ATOM, which explicitly incorporates the abstract syntax tree for representing code changes and integrates both retrieved and generated messages through hybrid ranking. Specifically, the hybrid ranking module can prioritize the most accurate message from both retrieved and generated messages regarding one code change. We evaluate the proposed model ATOM on our dataset crawled from 56 popular Java repositories. Experimental results demonstrate that ATOM increases the state-of-the-art models by 30.72\% in terms of BLEU-4 (an accuracy measure that is widely used to evaluate text generation systems). 
Qualitative analysis also demonstrates the effectiveness of ATOM in generating accurate code commit messages.

%  Commit messages document code changes (e.g., feature addition and bug repairs) in natural language, and are useful for program comprehension. However, due to the time cost and the lack of incentive, commit message writing is usually neglected by programmers, thus motivating us to automate the process. Previous studies on commit message generation have been benefited from generation-based models or retrieval-based models, but none of these studies consider integrating their advantages. For example, generation approaches can synthesize new commit messages, but cannot easily bridge the gap between code and natural languages; while the retrieval approaches cannot ensure the consistency of the identifiers. In this paper, we propose a novel retrieval-based commit message generation model to solve this issue. The proposed neural model explicitly exploits the retrieved candidates for enhancing the generator.
\end{abstract}

}

\maketitle
\IEEEdisplaynontitleabstractindextext
\IEEEpeerreviewmaketitle

\section{Introduction}
\IEEEPARstart{W}{ith} software growing in size and complexity, \sql{code hosting platforms}, {e.g.}, GitHub~\cite{Git} and TortoiseSVN~\cite{TortoiseSVN}, have been widely adopted in the life cycle of software development. These platforms greatly reduce time cost and maintenance cost. However, during the software updating, developers are required to submit commit messages to document code changes. The commit messages, which summarize what happened or explain why the changes were made, are usually described in natural language; thus the messages can help developers capture a high-level intuition without auditing implementation details. Hence, high-quality commit messages are essential for developers to comprehend version evolution rapidly. %\sen{[whether can help maintain the software?]}

However, manually writing commit messages is time-consuming and labour-intensive. First, until now, there is no specification regarding the writing format of commit messages when developers submit commits, and developers tend to follow their own writing styles. Second, developers tend to commit without writing the corresponding messages to make readers difficult to extract the precise description behind code changes manually. For example, according to the report ~\cite{dyer2013boa} in SourceForge~\cite{SourceForge}, an Open Source community dedicated to creating, collaborating and distributing projects, there are around 14\% of commit messages in more than 23,000 open-source Java projects \sql{that} are empty. \sq{Among} our collected dataset which contains the top-ranked $\sim$60 projects in terms of star numbers on GitHub, {e.g.}, Junit5~\cite{junit5} and Neo4j~\cite{neo4j}, we find that meaningless commit messages\footnote{Meaningless refers to empty, non-ASCII, merge and rollback commits.}
% Merge and rollback commits are removed as they often contain too many lines and these commits can be identified by the keywords "Merge", "Rollback".} 
also account for around 10\% of the entire collected commits. Therefore, automated generation of commit messages for code changes is necessitated and meaningful for software developers.
%\sen{Not sure whether we need to mention the quality of manual commit message. For example, some of them are incorrect in some ``poor'' projects.}
% , denoted as \texttt{diffs}, 

% \begin{lstlisting}[language=Java,caption={Example of the retrieved message by NNGen~\cite{DBLP:conf/kbse/LiuXHLXW18}, and generated messages by NTM~\cite{jiang2017towards} and the proposed ATOM for one code change.},label={list-1}]
% // Code changes
% -     clusterHealth = client("server1").admin().cluster().health(clusterHealthRequest().waitForGreenStatus()).actionGet();
% +     clusterHealth = client("server1").admin().cluster().health(clusterHealthRequest().waitForGreenStatus().waitForNodes("2")).actionGet();
% ------------------------------------------------
% // Reference Commit Message
%     "improve test to wait for NUMBER node"
% // Generated Commit Message by NNGen
%     "test wait until all machine join the cluster"
% // Generated Commit Message by NTM
%     "improve test to wait for NUMBER status"
% // Generated Commit Message by ATOM
%     "improve test to wait for NUMBER node"
% \end{lstlisting}

\begin{figure}[t]
     \centering
     \includegraphics[width=0.5\textwidth]{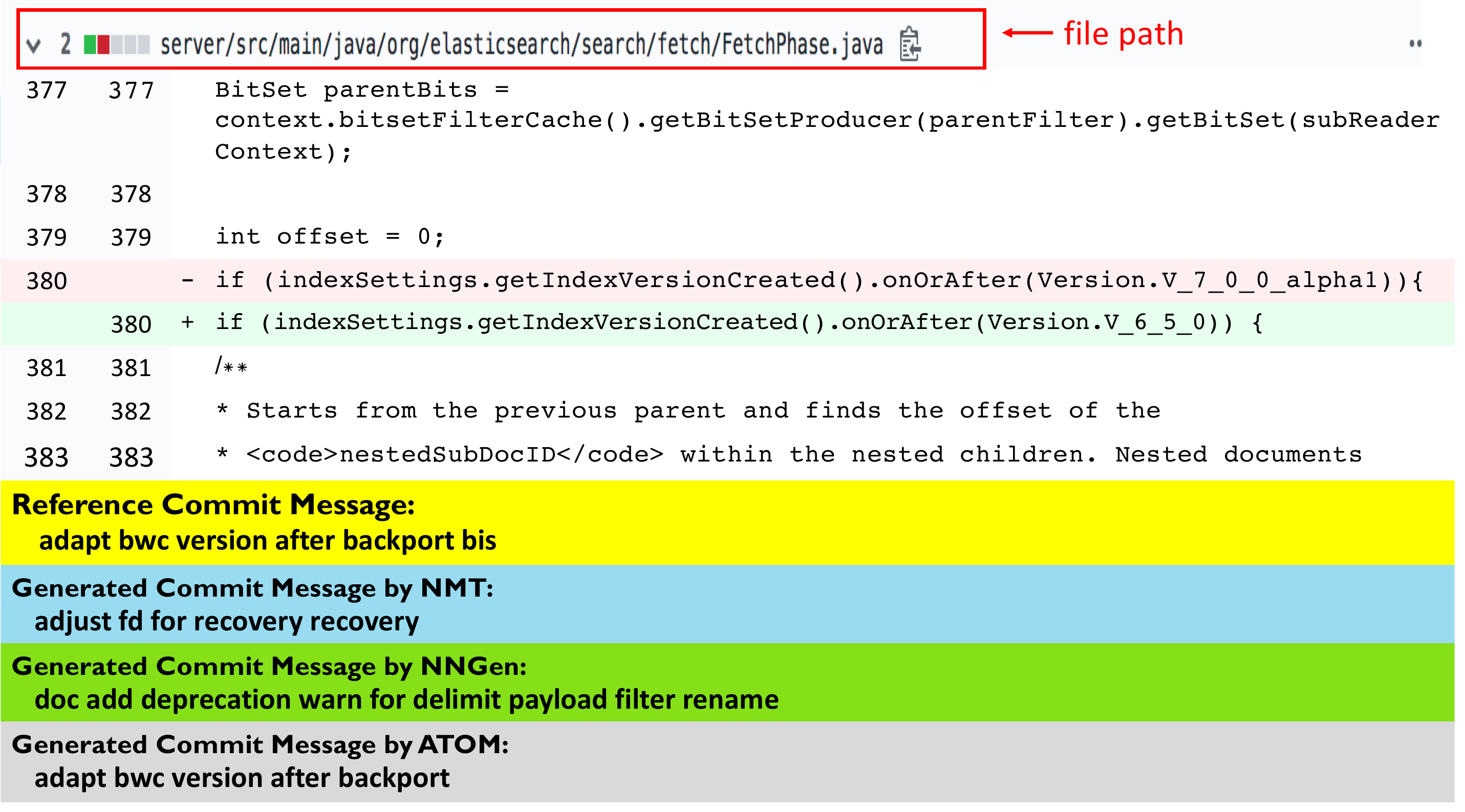}
     \caption{Example of the retrieved message by NNGen~\cite{DBLP:conf/kbse/LiuXHLXW18}, generated messages by NMT~\cite{DBLP:conf/kbse/JiangAM17}, and the proposed \tool for one code change of the commit \href{https://github.com/elastic/elasticsearch/commit/41528c0813fe72162408051e3af29ac42b4708f7}{\textit{41528c0813fe72162408051e3af29ac42b4708f7}}.} 
     \label{fig:example}
\end{figure}

\begin{figure}[t]
     \centering
     \includegraphics[width=0.5\textwidth]{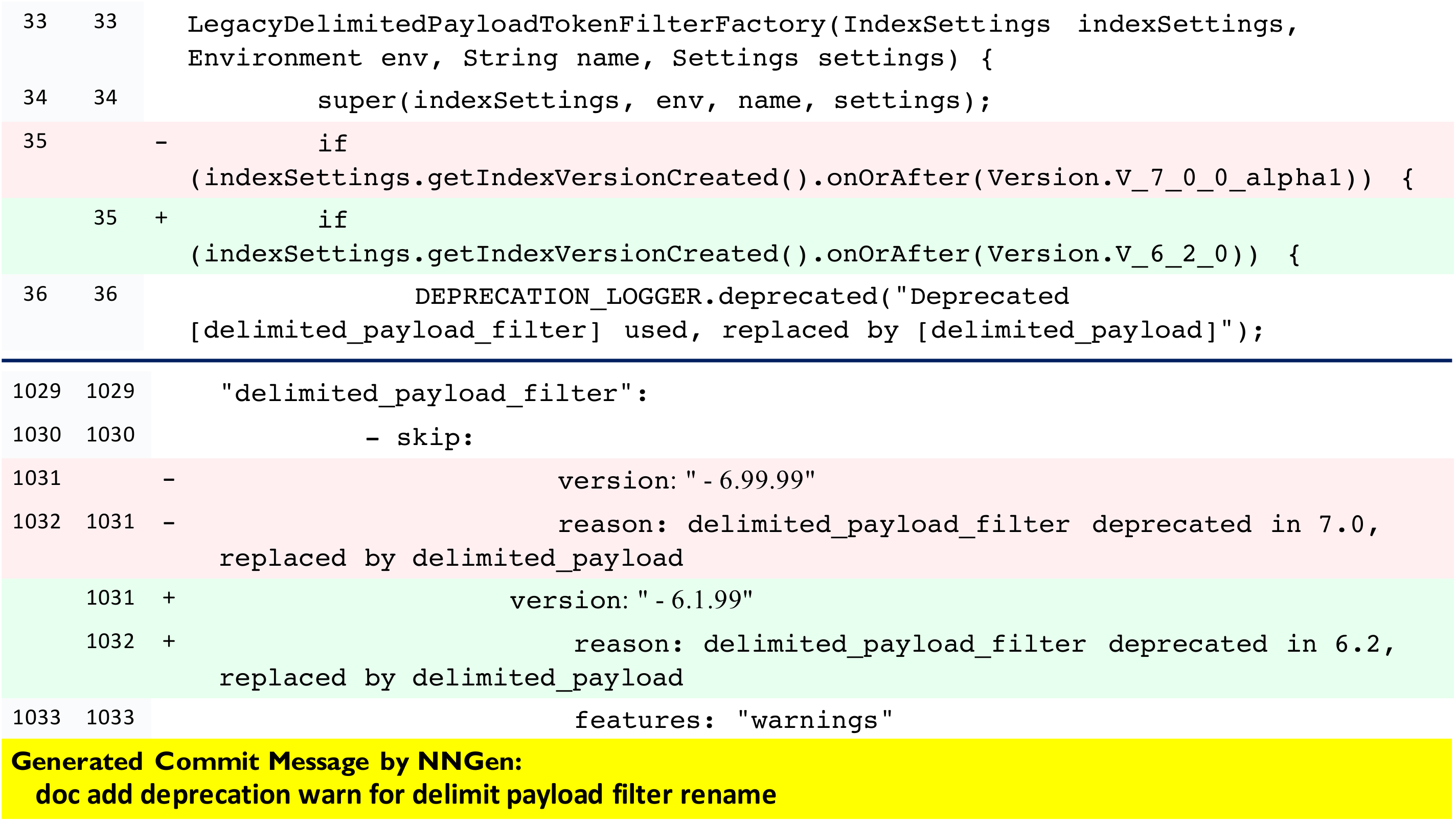}
     \caption{The code change of retrieved commit by NNGen~\cite{DBLP:conf/kbse/LiuXHLXW18} with its id  \href{https://github.com/elastic/elasticsearch/commit/c4fe7d3f7248223d5174b36fd4e1678217a6a6ed}{\textit{c4fe7d3f7248223d5174b36fd4e1678217a6a6ed}}.} 
     \label{fig:example_ret}
\end{figure}

Generating accurate commit messages by given code changes is a challenging task. Several approaches have been exhibited for generating commit message automatically.
The rule-based methods, {e.g.}, DeltaDoc \cite{DBLP:conf/kbse/BuseW10} and ChangeScribe \cite{ DBLP:conf/scam/Cortes-CoyVAP14}, are able to summarize code changes based on specific customized rules. However, these proposed rules could not easily cover all the cases and the generated messages are verbose, failing to capture the semantics behind a change~\cite{DBLP:conf/kbse/LiuXHLXW18}.
To deal with this limitation, Jiang et al. \cite{DBLP:conf/kbse/JiangAM17} proposed a generation-based approach, which adopts a neural machine translation (NMT) model for translating code changes into commit messages. However, the NMT model treats code as a flat sequence of tokens, which
ignores the syntactic and semantic information behind programs, thus fail to learn the semantics behind the code changes. Some other researchers \cite{DBLP:conf/esem/HuangZCXLL17,DBLP:conf/kbse/LiuXHLXW18} attempt to reuse the existing commit messages in the collected dataset by Information Retrieval to achieve the best performance. \sql{However, the retrieval-based approaches may achieve promising performance on similar programs, but are limited by the poorer performance on the programs that are very different from the retrieved database. For example, in Fig.~{\ref{fig:example}}, the message produced by retrieval-based approach, i.e., NNGen is unrelated to the code changes. Furthermore, Fig.~{\ref{fig:example_ret}} shows the retrieved commit of the commit in Fig.~{\ref{fig:example}}, where contains two parts, separating by a black line. 
%We can see that the first part is similar to the code changes in Fig.~{\ref{fig:example}}, but the retrieved message, e.g., \textit{delimit}, \textit{payload}, \textit{filter} are from the second part of code changes.
We can see that the first part is similar to the code changes in Fig.~{\ref{fig:example}}, but the retrieved messages (e.g., \textit{delimit}, \textit{payload}, and \textit{filter}) are from the second part of code changes.
Hence, the retrieval-based approach has no capacity to produce the exact commit messages on the dissimilar programs. Considering retrieval-based and generation-based techniques both have their merits, one intuition is to combine both for generating high-quality commit messages.}

\sql{To this end, we propose a novel commit message generation model, named {\tool} (\textbf{A}bstract syntax \textbf{T}ree-based c\textbf{O}mmit \textbf{M}essage generation) for better commit message generation. {\tool} contains three modules, 1) a generation module, which encodes the structure of changed code, i.e., Abstract Syntax Tree (AST), to enrich the semantic representation; 2) a retrieval module, which retrieves the most similar commit message based on the text-similarity; 3) a hybrid ranking module, which learns to prioritize the commit messages generated by generation and retrieval modules to further enhance the semantic relevance to the corresponding code changes.} To evaluate our proposed {\tool}, we crawl and build a new benchmark since AST cannot be constructed in the previous benchmarks~\cite{DBLP:conf/kbse/JiangAM17}. We quantitatively evaluate {\tool} on our crawled benchmark, including $\sim$160k commits in total. Extensive experimental results demonstrate that {\tool} can significantly outperform the state-of-the-art approaches by increasing at least 30.72\% in terms of BLEU-4 score~\cite{papineni2002bleu} (an accuracy measure that is widely used to evaluate text generation systems). Furthermore, {\tool} can enhance the performance of its generation module by 42.99\% by our well-designed hybrid ranking module.

The main contributions can be summarized as follows:
 \begin{itemize}
    \item We propose a novel generation module based on AST from code changes, named \textit{AST2seq}, to better capture code semantics and encode code changes.
    \item We design a hybrid ranking module to enhance the output of generation modules, by providing the most accurate commit messages among the generated and retrieved results.
    \item  We provide a new and well-cleaned benchmark, including complete function-level code snippets of $\sim$160k commits from 56 java projects. ~\liu {We clean the benchmark by filtering out meaningless (e.g., empty, non-ASCII, merge) commits and make the code ~\cite{Commit_code}
    % \footnote{\href{https://github.com/shangqing-liu/ATOM}{{\liu{https://github.com/shangqing-liu/ATOM}}}} 
    and benchmark~\cite{Commit_data}
    % \footnote{\href{https://drive.google.com/drive/folders/1Qw06U462e1nFUXEzXsclziMBpFrulykM?usp=sharing}{{https://drive.google.com/commit\_data}}} 
    public to benefit community research.}
    \item Extensive quantitative and qualitative experiments including a human evaluation demonstrate the effectiveness and usefulness of our proposed model.
\end{itemize}

\sql{The remainder of this paper is organized as follows. Section~\ref{sec:motivation} presents some basic knowledge about commits and neural networks. Then we describe the details about {\tool} in Section~\ref{sec:approach}. 
% and show experimental results and human study in Section~\ref{sec:expe}. 
{Experimental results, human evaluation, and \liu{examples} are conducted in Section~\ref{sec:expe}.}
% and followed by a case study in Section~\ref{sec:case}. 
Section~\ref{sec:discuss} gives some {discussion}
% discussions 
about {\tool}, followed by the related work in Section~\ref{sec:related}. Finally, Section~\ref{sec:conclusion} concludes this paper.}
\section{\sql{Motivation and} Background}\label{sec:motivation}
{In this section, we first introduce several features relevant to the commit\sql{, the motivation of our model design,} and some deep learning models/mechanisms used in our paper.}
%\sen{@shangqing, in this section, we may distinguish the background and related work.}

\begin{figure}[t]
     \centering
     \includegraphics[width=0.4\textwidth]{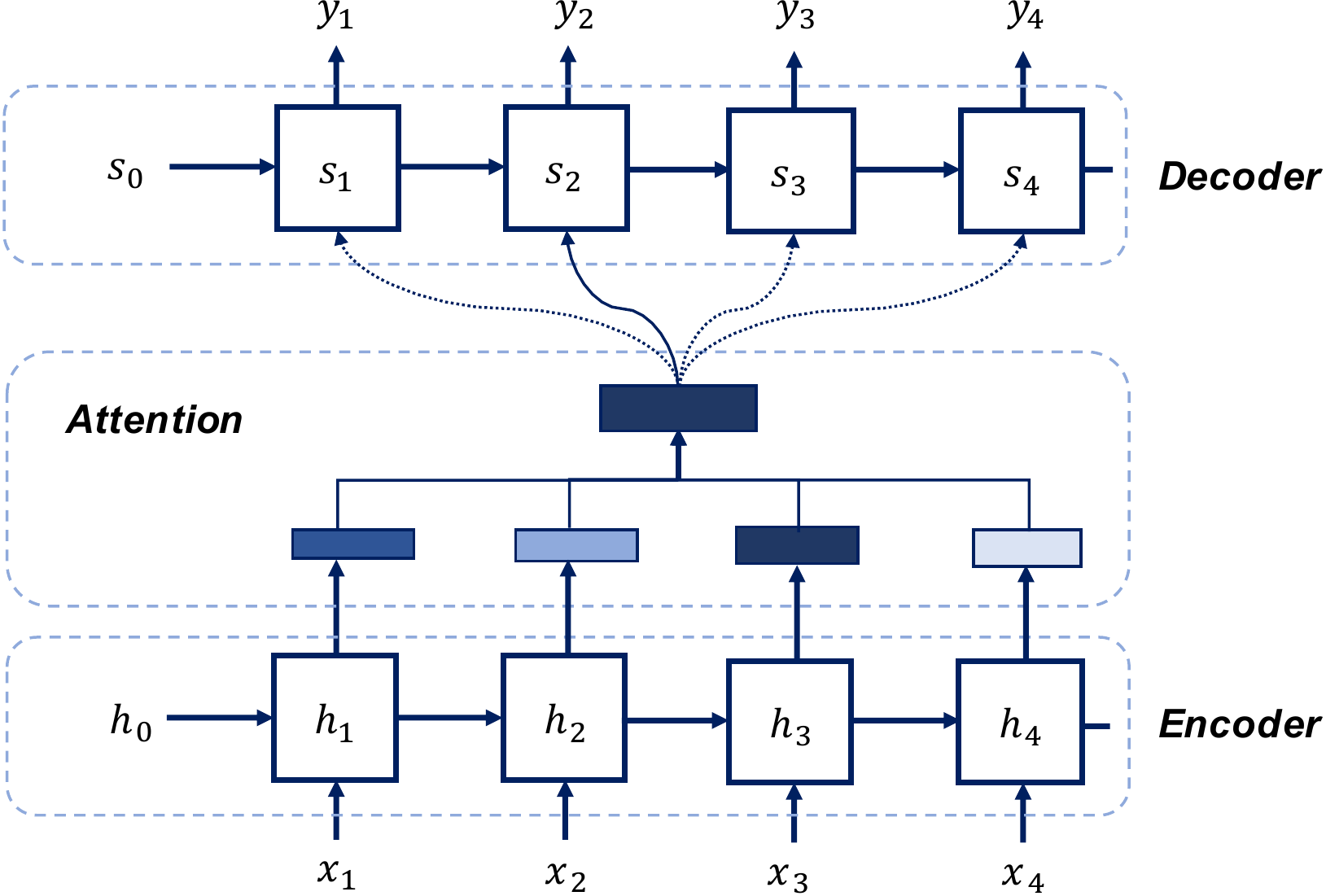}
     \caption{An encoder-decoder model with an attention mechanism.} 
     \label{fig:s2s}
\end{figure}

\subsection{Commit, \texttt{diff}, and Commit Messages}
%\sen{Code Change}
Commits are used in Git \cite{Git} to record the changes between different versions. As shown in Fig.~\ref{fig:example}, a commit usually contains a commit message and a change. The commit message is written by developers in a textual format to facilitate the understanding of current changes and the code change is called \texttt{diff} to characterize the difference between two code versions. Usually, a \texttt{diff} may contain one or multiple chunks with file paths, \sql{which can be found at a red rectangle in the upper part in Fig.~\ref{fig:example}}, along with the identifier ``diff --git`` to indicate the changed file name. The modified codes are wrapped by "@@" in a chunk with the negative sign '-' or positive sign '+' with a line number to denote the deleted or added line of code. Hence, we can summarize the commit in Fig.~\ref{fig:example}, in ``FetchPhase.java file``, there is one line of change at line number 380. We refer to the pair of \texttt{diff} and its corresponding message as a commit in this work.
%\sen{May provide the concrete scenario such as software development, and then introduce the specific definitions.}

\subsection{\sql{Motivation}}
\sql{Existing studies on commit message generation~\cite{DBLP:conf/kbse/JiangAM17,DBLP:conf/acl/LoyolaMM17, jiang2019boosting} generally treat the code changes as a sequence of code tokens and ignore the hierarchical code structure information. The work in other program comprehension tasks such as program vulnerability identification~\cite{zhou2019devign}, function name prediction~\cite{allamanis2017learning}, and source code summarization~\cite{fernandes2018structured,alon2018code2seq}, have utilized code structure such as Abstract Syntax Tree (AST) to learn code semantics and good performance has been demonstrated. Thus, in this paper\liu{,} we aim at exploiting AST for \liu{better-representing} code changes. Since code structures of code changes cannot be directly obtained by parsing functions, the usage of ASTs for the commit message generation task is more challenging. To well capture the semantics of long AST paths, we determine to use bi-directional \sq{LSTM}~\cite{gers1999learning} which shows effectiveness on representation learning of long-term sequences. However, a potential issue with bi-directional LSTM model is that the model needs to compress all the necessary information of the paths into a fixed-length vector~\cite{DBLP:conf/kbse/GaoZX0LK19}. To alleviate the issue, we follow the previous studies~\cite{alon2018code2seq, alon2019code2vec} to use the attention mechanism since attention can focus on some important paths to represent code changes.}

\subsection{Abstract Syntax Tree (AST)}
\begin{figure}[t]
     \centering
     \includegraphics[width=0.47\textwidth]{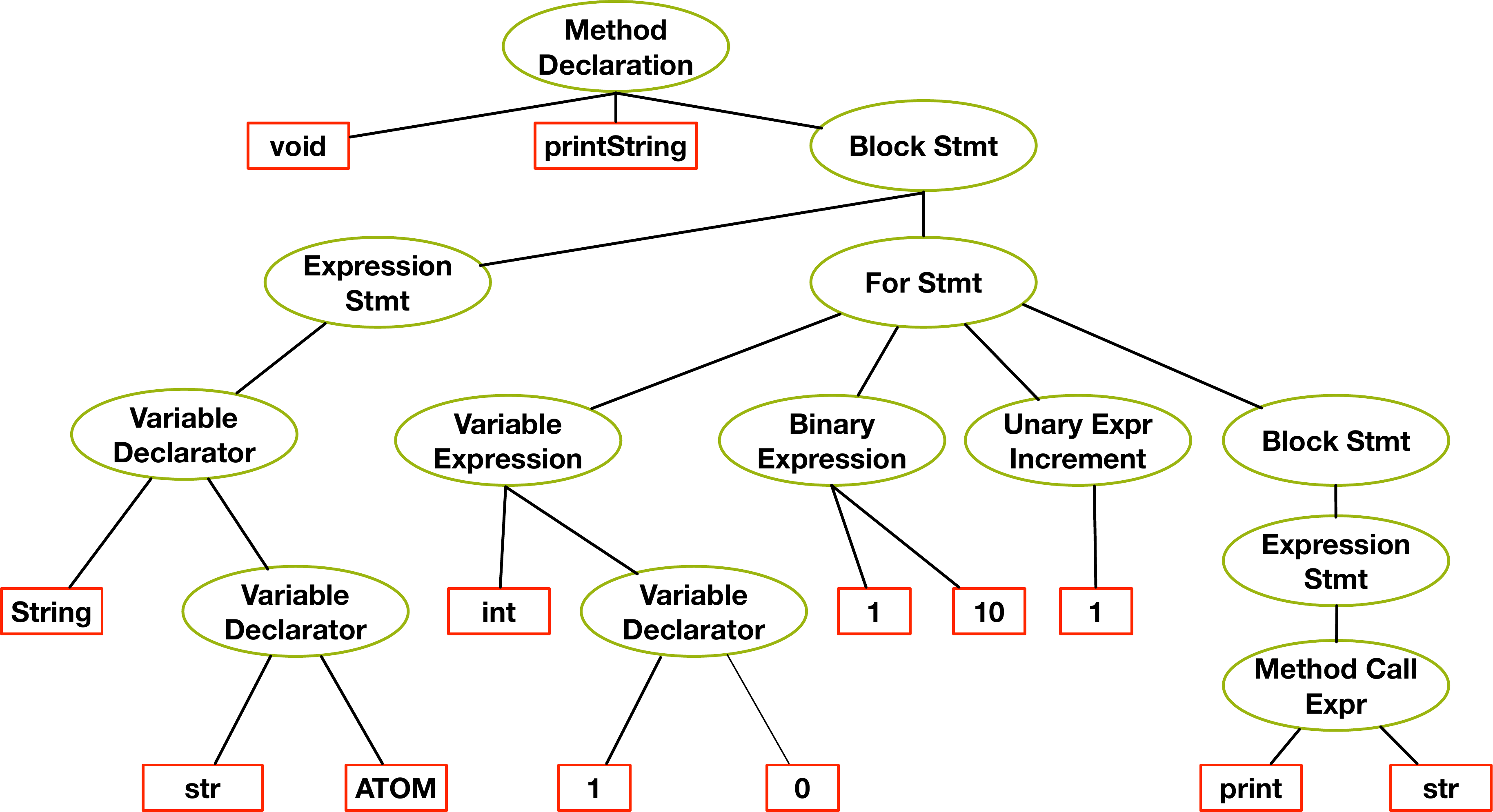}
     \caption{The AST compiled from Listing 1.} 
     \label{fig:ast}
\end{figure}

An abstract syntax tree is a high abstraction of source code, which is a tree structure \sql{and serves} as the intermediate representation of program language. An AST usually contains leaf nodes that represent identifier and literal in the code and non-leaf nodes which can represent some syntactic \sql{structure} within codes. Specifically, Fig.~\ref{fig:ast} shows a simple AST with the code snippet in Listing~\ref{code_snippet}, where identifier name e.g.  ``str``, ``ATOM`` or type e.g. ``int``, ``String`` are represented by the values of leaf nodes and  non-leaf nodes e.g. ``ExpressionStmt``, ``ForStmt`` tend to have more syntactic information. We can get a total of 106 different non-leaf nodes with \textit{JavaParser}~\cite{javaparser}, which is a tool used for extracting ASTs in Java language. Adopting AST in code comprehension has been proved to get the state-of-the-art performance, such as code2seq \cite{alon2018code2seq}, code2vec \cite{alon2019code2vec}, DeepCom \cite{hu2018deep}, CRF \cite{alon2018general}, Devign \cite{zhou2019devign}.

\begin{lstlisting}[style=mystyle, caption=A simple Java code snippet, float=tp, label={code_snippet}]
public void printString(){
    String str = "ATOM"
    for(int i = 0; i < 10; i++){
        print(str);
    }
}
\end{lstlisting}

\subsection{Encoder-Decoder Model}\label{sec:en-de}
The basic structure of NMT \cite{wu2016google} used to translate source sequences into targets is encoder-decoder, as shown in Fig.~\ref{fig:s2s}. The feature vectors generated by encoder are fed into the decoder to generate target sequences. Usually, it consists \sql{of} two RNNs \cite{cho2014learning} with built-in LSTM cells \cite{hochreiter1997long} and attention mechanism \cite{bahdanau2014neural} \cite{luong2015effective} for translation.
\subsubsection{Recurrent Neural Network (RNN)}
RNNs are widely used to capture information from time-series data as their chain-like natures. The loop \sql{contained} in RNNs allows information to be passed from \liu{one-time} step to the next. At each time $t$, the unit in RNNs takes $x_t$ and the hidden state $h_{t-1}$, which is produced by previous time $t-1$ as input to predict the current output $y_t$. The chain-like structure enables RNNs to learn information from the past, however\liu{,} they also suffer from long-term dependencies. Since RNNs are unable to connect information from further back and cannot handle long sequences, some variants e.g., Long Short-term Memory (LSTM) \cite{hochreiter1997long} and Gated Recurrent Unit (GRU) \cite{chung2014empirical} are proposed.
\subsubsection{Long Short-Term Memory (LSTM)}
LSTMs introduced by \sql{Hochreiter and Schmidhube~\cite{hochreiter1997long}} are explicitly designed with a memory cell to remember important information. The gating mechanism in the memory cell helps LSTMs selectively `forget' unimportant information, thus allowing more space to take in information and controls when and how to read previous information and write new information. In this way, the memory cell will preserve \sql{more} long-term dependencies than vanilla RNNs. Hence, RNNs built with LSTMs are widely used for sequence models to capture information.
%\sen{Please provide some relations between our topic and LSTM.}

\subsubsection{Attention Mechanism}
%\sen{We forget mention attention mechanism in the intro?}
Attention is proposed to boost the performance of Encoder-Decoder further, as it utilizes all the hidden states of the input sequence rather than the final hidden state as a context vector for the decoder. It creates an attention mapping matrix between each time step of the decoder output to the encoder hidden states. The attention weights are trained by a forward neural network to align the scores between the encoder states and the decoder outputs. This means, for each output, it has access to the entire input sequence and dynamically selects specific elements from the input. Hence, the attention mechanism allows the decoder to focus and place more \textit{Attention} on the relevant parts of the input. The Bahdanau \cite{bahdanau2014neural} or Luong \cite{luong2015effective} Attention has been widely adopted into neural machine translation \cite{vaswani2017attention}, reading comprehension \cite{levy2017zero} and computer vision \cite{ba2014multiple}.

\subsection{Convolutional Neural Network (CNN)}
Consisting of Convolutional Layer, Pooling Layer and Fully-Connected Layer, Convolutional Neural Networks (CNNs) are one of the most common Deep Neural Network \sql{architectures}. The convolution operations apply kernels to extract features from the feature maps, which allow the network to capture \liu{high-level} abstract information with a reduced number of parameters. In image processing tasks, for instance, the convolution layers can learn edges, patterns and shapes after training. Similarly, CNN can also be applied to natural language processing tasks. 
\sql{In previous work \cite{hu2014convolutional, pang2016text}}, CNN-based neural ranking models are trained to learn \liu{high-level} sentence matching patterns. 
%\section{Retrieval-Based Commit Message Generation Model \sen{update}}\label{sec:approach}
\section{Our approach (\tool)}\label{sec:approach}
\begin{figure}[t]
     \centering
     \includegraphics[width=0.49\textwidth]{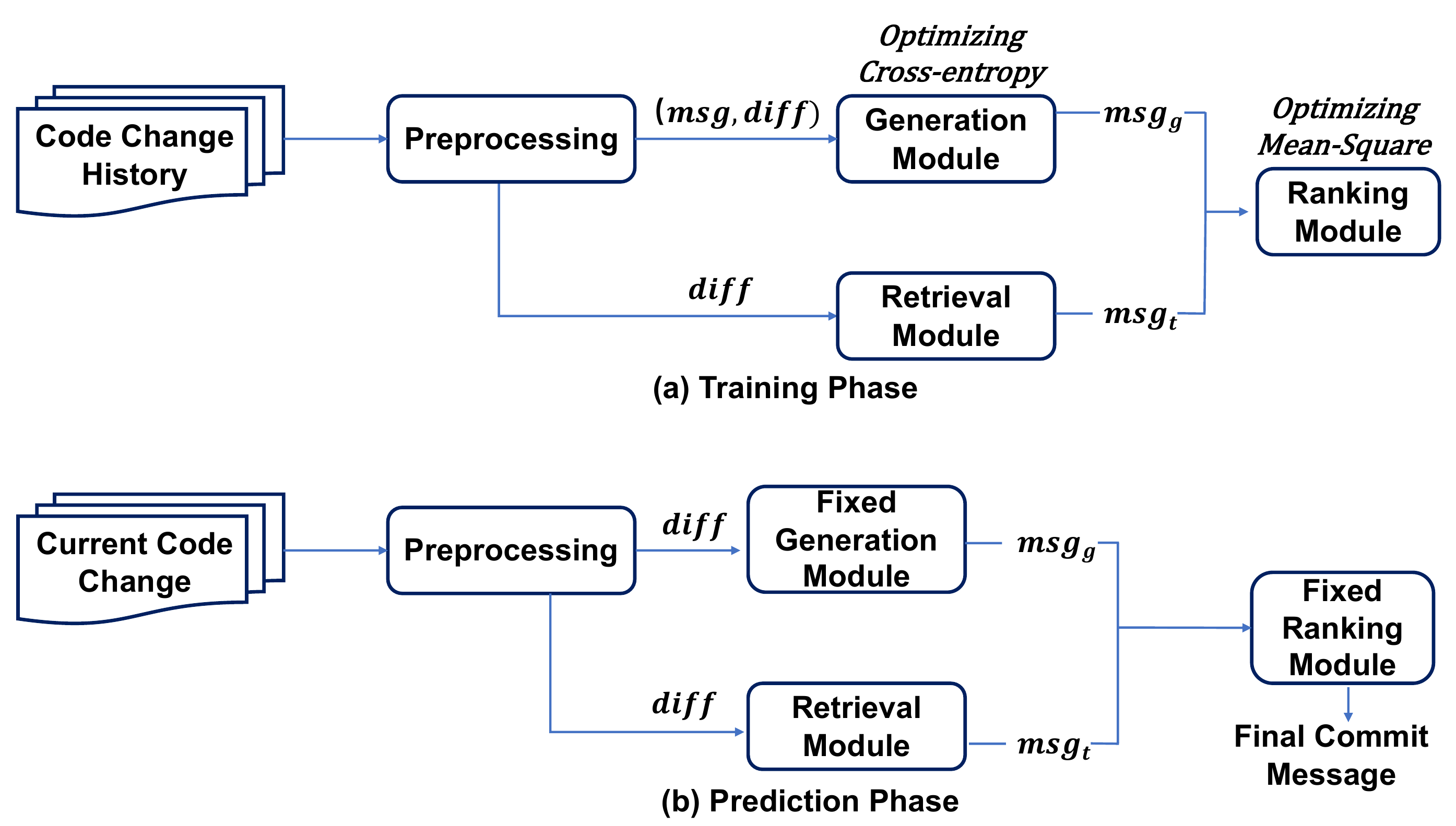}
     \caption{\sq{Architecture of \tool.}}
     \label{fig:framework}
\end{figure}

\sql{In this section, we provide the overview of our approach {\tool} and detail each of the modules.}

% first briefly introduce the motivation behind the model design, then

\subsection{Overview}
Fig.~\ref{fig:framework} shows the architecture of our framework, which consists the following components.
 \begin{itemize}
    \item \textit{Preprocessing Module.} The commit message and code changes are processed separately. We extract AST paths corresponding to the code changes by retrieving the completed functions in the repository. We also use the first sentence with lemmatization from the commit message as the target sentence to represent the entire commit message.
    \item \textit{Generation Module.} We name our generation module as \textit{AST2seq} and encode AST paths from \texttt{diff}s with BiLSTM to represent code changes and followed a decoder with \liu{an} attention mechanism to generate a new message $msg_g$.
    \item \textit{Retrieval Module.} The retrieval module uses a ``\texttt{diff}-\texttt{diff}`` match to retrieve the most relevant commit messages. This approach matches \texttt{diff} with all \texttt{diff}s in the training set and get the most relevant message $msg_t$ based on the cosine similarity.
    \item  \textit{Ranking Module.} To incorporate the retrieval results into the generation module, we train a CNN to adaptively rank the generated message $msg_g$ and the retrieved message $msg_t$.
\end{itemize}
\sq{At the prediction phase, when a new code change arrives, \tool generates the corresponding commit message with the trained generation module and ranking module, as shown in Figure~\ref{fig:framework} (b).}

\subsection{Preprocessing Module} 
We preprocess code changes and commit messages separately for preparing the input of \tool.

\subsubsection{Code Changes} 
We first divide code changes \texttt{diff}s into \textit{added} and \textit{deleted} groups based on the corresponding sign, i.e., ``+'' and ``-''. Then we tokenize the \texttt{diff}s with pygments~\cite{pygments}, and remove meaningless tokens such as punctuations. %After such step, 
Consequently,
we obtain a list of tokens for the \textit{added} code and \textit{deleted} code, denoted as $W^+$ and $W^-$ respectively, where $W^{+/-} = \{w_1, w_2, ..., w_i\}$ and $i$ is the $i$-th token in the changed code.

% Since we need to extract AST paths corresponding to \texttt{diff}s based on the basic compilation unit, which contains a single class definition and wrapped functions. 
\sq{The basic compilation unit~\cite{niemeyer2005learning}, containing a single class definition and wrapped functions, is needed to extract AST paths based on the \texttt{diff}s.}
Hence, we retrieve completed functions of \texttt{diff}s denoted as \textit{added} function and \textit{deleted} function. We use Ctags~\cite{ctags} with file paths and modified line numbers containing in \texttt{diff}s to retrieve completed functions in the repository and then parse them to obtain ASTs with JavaParser~\cite{javaparser}. As all tokens belonging to $W^{+/-}$ are the values of leave nodes in an AST, we search the shortest distance \footnote{Here the shortest distance refers to the minimum edges between two corresponding leave nodes.} for any two tokens, $w_i$ and $w_j$ in $W^{+/-}$ and denote the path as $x = \{w_i, n_1, ..., n_l, w_j\}$, where $n_l$ means the $l$-th non-leaf node. Following this procedure, we finally obtain AST paths for the whole \textit{added}/\textit{deleted} code, indicated as a set of $X^{+/-}=\{x^{+/-}_1, ..., x^{+/-}_{p/k} \}$ where $p$, $k$ are the total number of AST paths respectively.

\subsubsection{Commit Messages} 
We extract the first sentence from the commit message as the target sequence since the first sentence is often the \sq{summary} of the entire commit~\cite{jiang2017towards}\cite{DBLP:conf/kbse/JiangAM17}\cite{gu2016deep}. We split the tokens with underlines ``$\_$'' and replace file names and digits with unique placeholders ``\textless FILE\textgreater'' and ``\textless NUMBER\textgreater'' respectively. We also lemmatize each word into its base form by using the NLTK toolkit~\cite{nltk} to reduce the vocabulary set. The lemmatized message is denoted as $M=\{y_1, y_2, ..., y_n\}$ where $n$ is the token length of the message.

\begin{figure*}[t]
     \centering
     \begin{subfigure}[h]{0.48\textwidth}
        \centering
    	\includegraphics[width=1 \textwidth]{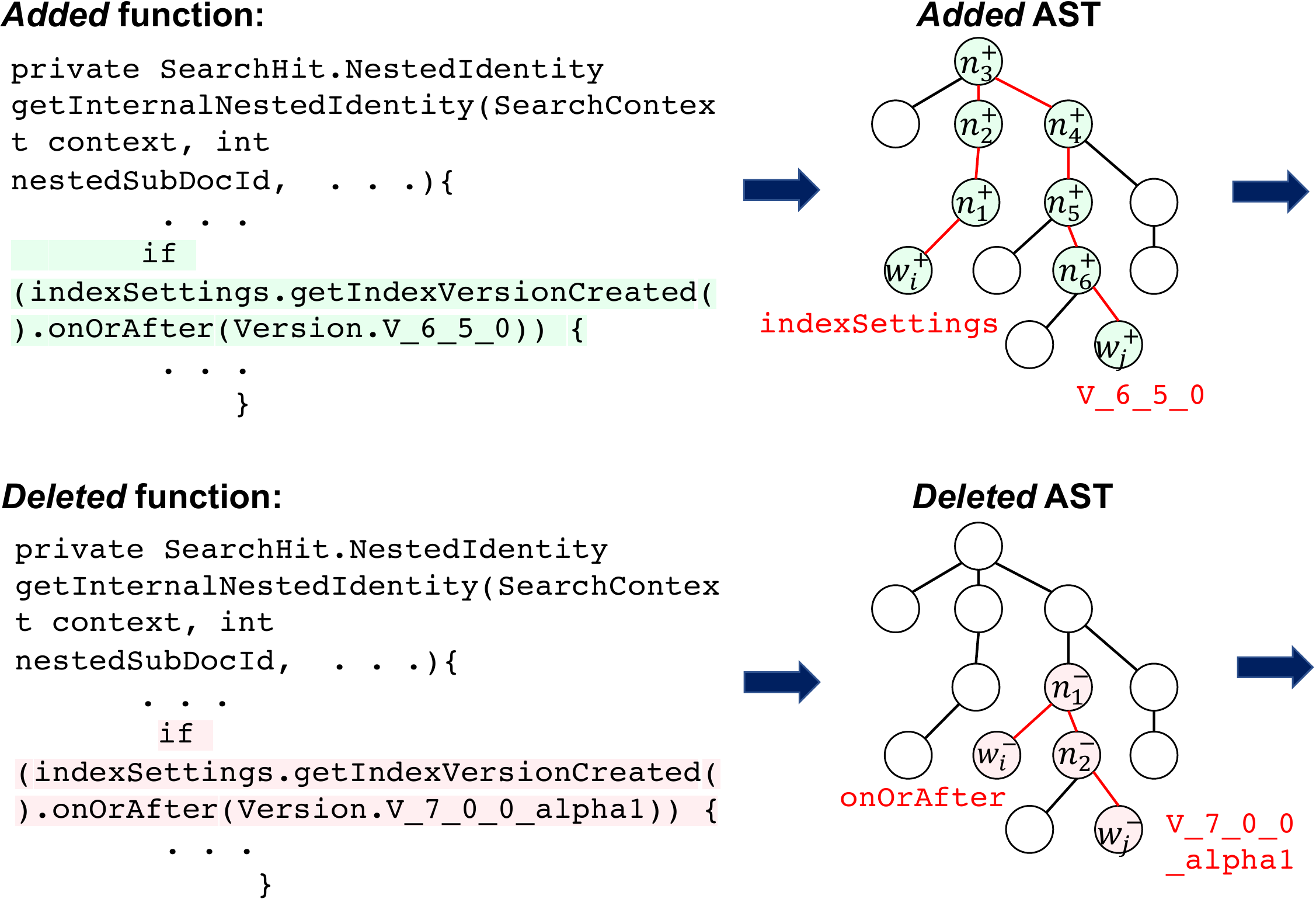}
        \end{subfigure}
        \begin{subfigure}[h]{0.45\textwidth}
        \centering
        \includegraphics[width=0.95 \textwidth]{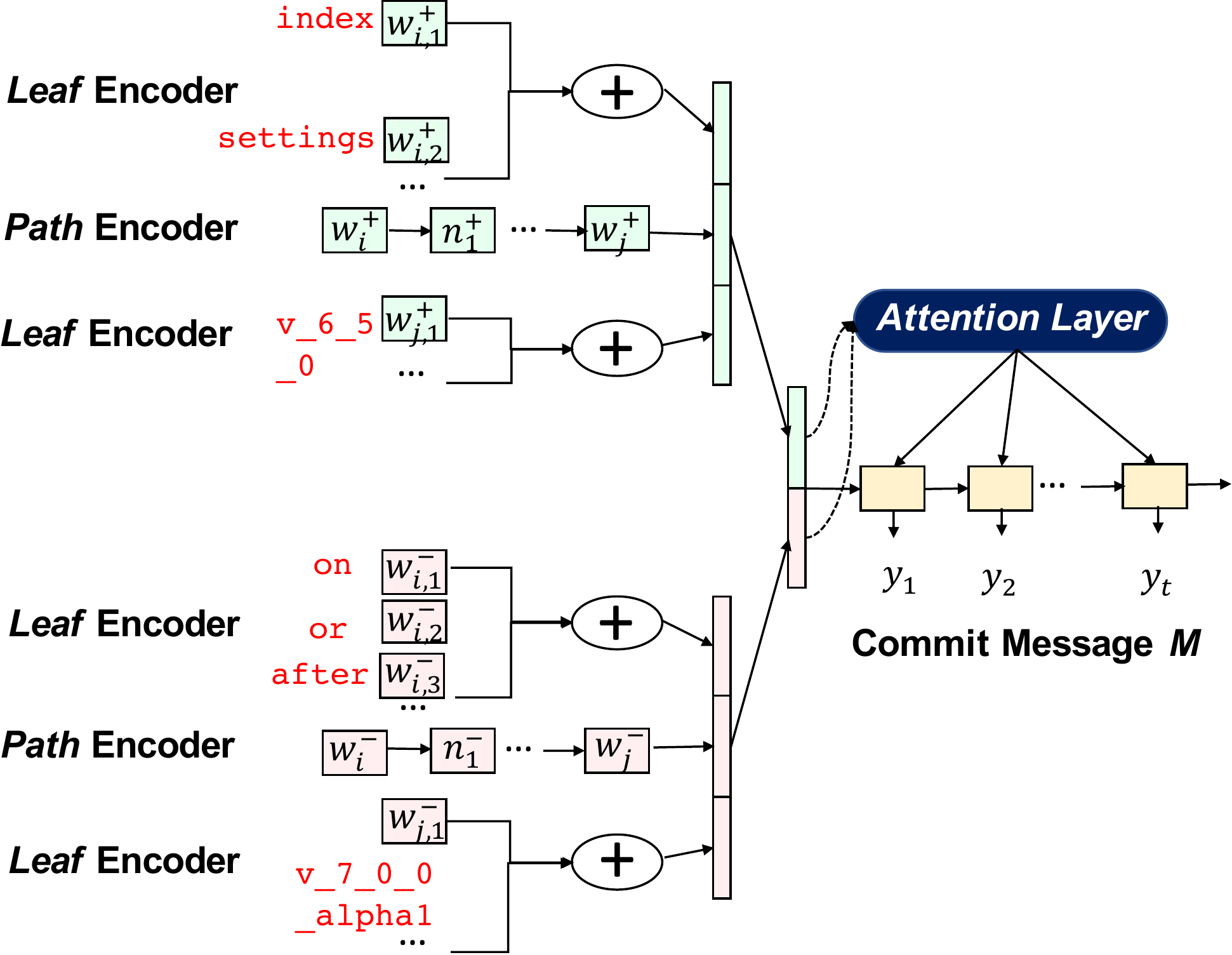}
        \end{subfigure}
     \caption{Architecture of \textit{AST2seq} with the example in Fig.~\ref{fig:example}, where \textit{added} and \textit{deleted} function denote the completed functions retrieved from the \texttt{diff}. The highlight path in \textit{added} or \textit{deleted} AST is one of paths extracted from tokens, e.g., indexSettings, onOrAfter in \texttt{diff} and $n_l^{+/-}$ is the non-leaf node, e.g., ``ForStmt``, ``Binary Expression``.}
     \label{fig:ast2seq}
\end{figure*}

\subsection{Generation Module}\label{sec:gen}
Prior work on commit message generation treated \texttt{diff}s as a flat sequence of tokens, which is limited by long sequences and \sq{ignores the code structure, e.g., Abstract Syntax Trees (ASTs) of the \texttt{diff}s, to capture the semantics.} AST is an abstraction of code and has been proved useful in representing code semantics~\cite{alon2018code2seq}\cite{alon2019code2vec}\cite{hu2018deep}. However, the AST-based approaches mostly extract ASTs on the completed functions to understand the functionality of codes. In our generation module, we extract the AST paths based on the \texttt{diff}s for representing code changes.
Compared to the sequence-based approaches, our method can generate messages with longer \texttt{diff}s and we name as \subtool. The entire architecture of \subtool is illustrated in Fig. \ref{fig:ast2seq}, involving three main components sequentially: \textit{AST Encoder} for encoding each AST path into its vector representation; \textit{Attention} for dynamically focusing on the relevant AST paths; and \textit{Message Decoder} for generating corresponding commit message of the code change.

% Encoding an Abstract Syntax Tree (AST) to represent a source code snippet has been proved to get an start-of-the-art performance on code summarization \cite{alon2018code2seq}. Inspired by this, we designed our Generation Module \textit{Ast2seq} to generate message $msg_g$ automatically. \textit{Ast2seq} architecture shown in Figure \ref{fig:ast2seq}, which includes three sequential components: \textit{Diff Ast Encoder}, which encodes each AST path into a vector representation, \textit{Attention}, which dynamically select the distribution over AST paths, \textit{Decoder}, which generated message to describe the code changes.

\subsubsection{AST Encoder}
Given a set of \textit{added} and \textit{deleted} AST paths $X=\{x^{+/-}_1, x^{+/-}_2, ..., x^{+/-}_{p/k}\}$, where $x \in X$ can be represented as $\{w_i,n_1,...,n_l,w_j\}$ and $p, k$ are the \textit{added} and \textit{deleted} AST paths. 
We encode each path $x$ with a bi-directional LSTM to create a vector representation $z$. \sql{Here we use bi-directional LSTM is to expect the bi-directional LSTM can capture the long-term semantics in each AST path}.

\begin{itemize}
    \item \textbf{Path Representation.} The types of nodes e.g., ``ForStmt``, ``IfStmt`` that make up an AST path $x$ is limited to 106 and we represent these node types with an embedding matrix $E^{nodes}$ and then encode the path e.g., $\{w_i^-, n_1^-, n_2^-, w_j^-\}$ in Fig.~\ref{fig:ast2seq} into a bi-directional LSTM to obtain the dense representation $ h_{w_i},...,h_{w_j} $ and use the final states of LSTM as path representation.
    % to get the final states. We can formulate this in the following:
    \begin{equation}
        h_{w^{+}_i},...,h_{w^{+}_j} = LSTM(E^{nodes}_{w^{+}_i},...,E^{nodes}_{w^{+}_j})
    \end{equation}
    \begin{equation}
        path\_feat^{+} = [h^\leftarrow_{w^{+}_i};h^\rightarrow_{w^{+}_j}]
    \end{equation}
    \begin{equation}
        h_{w^{-}_i},...,h_{w^{-}_j} = LSTM(E^{nodes}_{w^{-}_i},...,E^{nodes}_{w^{-}_j})
    \end{equation}
    \begin{equation}
        path\_feat^{-}= [h^\leftarrow_{w^{-}_i};h^\rightarrow_{w^{-}_j}]
    \end{equation}
    \item \textbf{Leaf Representation.} As the values of start leaf node $w_i$ and end leaf node $w_j$ of an AST path also appear in the \texttt{diff}, we incorporate them for representing a completed path. We split the tokens of the values in leaf nodes e.g., ``onOrAfter`` in Fig.~\ref{fig:ast2seq} into subtokens, ``on``, ``or``, ``after`` and then combine the embeddings of these subtokens with summation to represent a leaf token:
    
    % are leaf nodes whose values are also in the diff. Hence, we split tokens into subtokens and sum the subtoken vectors with an embedding matrix $E^{subtokens}$ to represent the leaf tokens:
    \begin{equation}
        leaf_\_feat_{w^{+}} = \sum_{s \in split(w^{+})} E^{subtokens}_{w^{+}}[s]
    \end{equation}
    \begin{equation}
        leaf_\_feat_{w^{-}} = \sum_{s \in split(w^{-})} E^{subtokens}_{w^{-}}[s]
    \end{equation}
\end{itemize}

To represent a completed path $x^{+/-}$, we aggregate the path representation and leaf representation by employing a fully connected layer:
\begin{equation}
\small
        z^{+/-} = layer([leaf_\_feat_{w^{+/-}_i}; path\_feat^{+/-}; leaf_\_feat_{w^{+/-}_j}])
\end{equation}
Finally, we concatenate $p$ \textit{added} and $k$ \textit{deleted} paths of vector $z$ for representing a \texttt{diff}:
% the changed feature vectors:
\begin{equation}
        Z = [z^+_p;z^-_k]
\end{equation}

\subsubsection{\textit{Attention}}
\sql{Given a set of \textit{added} and \textit{deleted} AST path representations $Z=\{z_1,z_2,...,z_{p+k}\}$, where $p+k$ is the summation of \textit{added} and \textit{deleted} paths, we need to focus on some important paths which can capture the information to represent the entire code changes. Hence, attention is needed to learns how much focus \sq{``attention'' should} be given to each AST path. We use Luong Attention Mechanism \cite{luong2015effective}, which is shown in the equation~\ref{luong}. During decoding period, the attention will learn the weight distribution over these paths to capture the important paths.}

\subsubsection{Message Decoder}
We take the average of vector representations of \textit{added} and \textit{deleted} paths, i.e., $Z=\{z_1,z_2,...,z_{p+k}\}$, as an initial hidden state of the decoder, that is:
\begin{equation}
        h_0 = \frac{1}{p+k}\sum_{i=1}^{p+k} z_{i}.
\end{equation}
At each decoding step $t$, a context vector $c_t$ is computed based on $Z$ and current hidden state $h_t$ in the decoder.
% Here $Z$ is analogous to source-side context vector in language translation:
\begin{equation}
\label{luong}
       \alpha^t = softmax(h_t W_\alpha Z),   \qquad  c_t = \sum_{i}^{p+k} \alpha_i^t z_i.
\end{equation}
$\alpha^t$ is the variable-length alignment vector whose size equals the number of \textit{added} and \textit{deleted} paths. 
Then $c_t$ and $h_t$ are combined to predict the current token $y_t$ \cite{luong2015effective}:
\begin{equation}
    p(y_t|y<t, z_1,...,z_{p+k})=softmax(W_s tanh(W_c[c_t;h_t])) 
\end{equation}
The loss function we adopted in \subtool is softmax cross entropy with logits:
\begin{equation}
    Loss = y \log(\frac{e^{logits}}{\sum_r e^{logits}}).
\end{equation}
where $r$ is the commit message vocabulary size, $y$ is the true token of message and $logits$ is the output of decoder module.

\begin{figure}[t]
     \centering
     \includegraphics[width=0.48\textwidth]{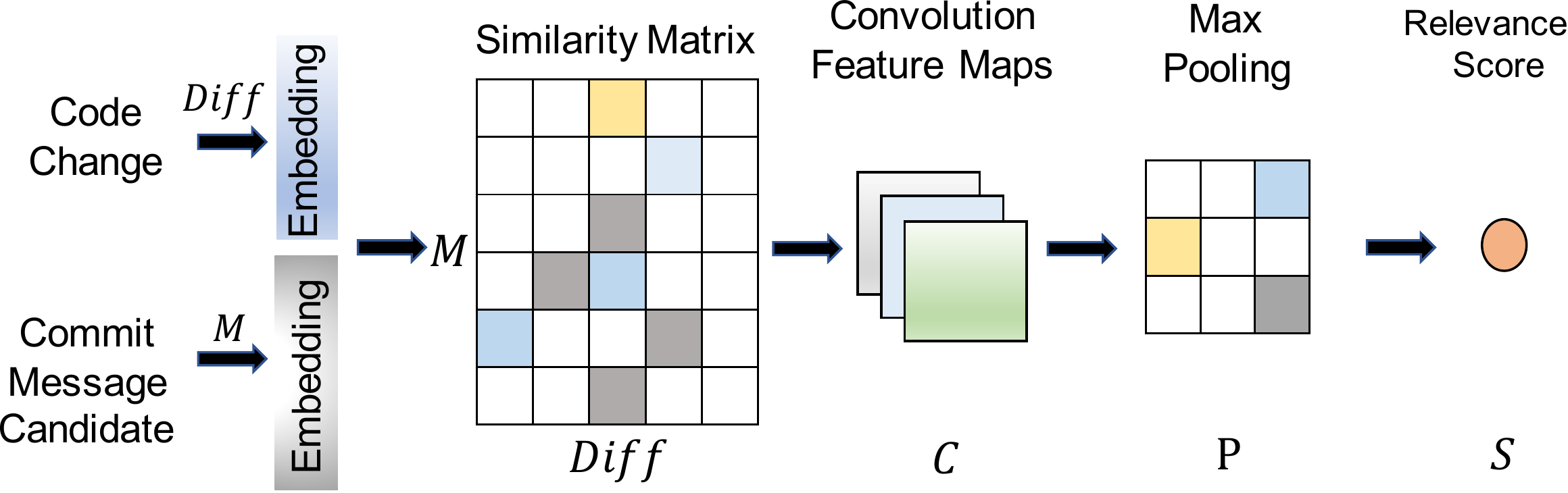}
     \caption{Architecture of ConvNet.} 
     \label{fig:convnet}
\end{figure}

\subsection{Retrieval Module}\label{sec:ret}
The retrieval module aims at retrieving relevant commit messages from the training set. We adopt a ``\texttt{diff}-\texttt{diff}'' match for the retrieval. Specifically, we first index all \texttt{diff}s in training set with sklearn~\cite{sklearn}. Then for each \texttt{diff} in the validation and test sets, we compute the cosine similarity in the training set based on their tokens of tf-idf scores \cite{aizawa2003information}, and keep the most relevant one commit message (first-ranked) from the training set. Term frequency (TF) and inverse document frequency (IDF) can be computed by the following equation:

\begin{align}
  tf_{i,d} = \frac{n_{i,d}}{\sum_{i \in {W}} n_{i,d}} && idf(i) = log(\frac{N_{diff}}{df_i})
\end{align}
where $n_{i,d}$ is the number of $i_{th}$ token in the $d$ and $W$ is the set of distinctive tokens. In the second equation, $df_i$ is the number of \texttt{diff}s that contains $i_{th}$ token in the entire \texttt{diff}s and $N_{diff}$ is the total number of \texttt{diff}s.

The retrieved commit message serves as one candidate for the final generated message, and will be fed into the ranking module together with the message produced by the generation module.

\subsection{Hybrid Ranking Module}
\sql{From the retrieval module (described in Section 3.5) and generation module (described in Section 3.4), we can get two commit message candidates $y \in \{msg_t, msg_g\}$ where $msg_t$ is the first-ranked retrieved commit message and $msg_g$ is the generated message. To predict which candidate is better, we can train a binary classifier based on popular models such as XGBoost~\cite{chen2016xgboost} and LSTM~\cite{hochreiter1997long}. However, since \texttt{diff}s may contain tokens that tend to appear in the generated messages, e.g., tokens related to function name and variable name, the relevance between \texttt{diff}s and candidate messages would be useful for the final message prediction. Inspired by Liu et al.~\cite{yang2019hybrid}, we design a similarity matching matrix to measure the relevance between \texttt{diff} and the corresponding candidate message, and adopt ConvNet model to learn their relevance score. Experiments in Section 4.3.4 show that ConvNet outperforms typical classifiers (e.g.,  XGBoost~\cite{chen2016xgboost} and LSTM~\cite{hochreiter1997long}).}

\subsubsection{Similarity Matching Matrix } \label{sec:matrix}
% Interaction
For any \texttt{diff} $d$, we first looks up embeddings for tokens in $d$ and $y$ respectively, denoted as $E(d) = [d_1, d_2,...,d_{L_d}]$ and $E(y) = [y_1, y_2,...,y_{L_y}]$ where $L_d$ and $L_y$ are the lengths of $d$ and $y$ respectively.
Note that the embedding matrixes for \texttt{diff}s and messages are trained separately, but their dimension sizes are equal. The interaction matching
matrix $D$ is computed by the following equation:
\begin{equation}
    D = E(d) \times E^\top(y),
\end{equation}
where $D$ has the dimension with $(L_d, L_y)$ and is used as the input of ConvNet to predict a matching score.

\subsubsection{ConvNet Model} \label{sec:convnet}
ConvNet is designed to find the correlation between \texttt{diff} and message and give a better output among commit message candidates which contains convolution and max-pooling operations on the similarity matching matrix $D$. Let $C_{in}$ denote \liu{the number} of input channels, $H$ is the height of input plane and $W$ is width, which in our initial settings equal to $1$, $L_y$ and $L_d$ respectively.
% Hence, 
The convolution operation $out(C_{out})$ on the input $input$ with size $(C_{in}, H, W)$ and output size $(C_{out}, H_{out}, W_{out})$ can be expressed as:
\begin{equation}
    out(C_{out}) = \sigma(bias(C_{out}) + \sum_{k=0}^{C_{in}-1}weight(C_{out}, k) * input(k)
\end{equation}
where $\sigma$ is the activation function, and $*$ is the valid dot product operator. 
Max-pooling operation with input size $(C,H,W)$ is conducted after the convolution operation, which can be expressed as:
\begin{equation}
\begin{split}
out(C, h, w) = \max_{m=0,...,kH-1} \max_{n=0,...,kW-1}\\ 
      input(C, stride[0] \times h + m, stride[1] \times w + n)
\end{split}
\end{equation}
where $(kH, kW)$ is the kernel size and $stride[\cdot]$ is the tuple of the sliding blocks over the input, $stride[0]$ and $stride[1]$ represent the block height and width correspondingly. Finally we feed the output produced by ConvNet into a fully connected layer to compute the relevance score between \texttt{diff} $d$ and message $y$. We use Mean Square Error Loss function~\cite{james1992estimation} to optimize the loss values in the form of:

\begin{equation}
    Loss = |Y^\prime-Y|^2
\end{equation}
where $Y^\prime$ is the output of 
% our 
ConvNet and Y is true relevance score.

\subsubsection{Training For ConvNet}
One challenging part in the hybrid ranking module is how to well define the true relevance scores $Y$ between \texttt{diff}s and corresponding candidate messages. One possible solution is to manually evaluate these candidates, however, the time and labour cost would be very intensive and it is not applicable for end-to-end training. To enable an end-to-end training process, we propose to build upon the evaluation metrics, e.g., BLEU-4~\cite{yang2019hybrid, song2018ensemble}.
Specifically, we score these two candidate messages by comparing them with the ground truth using BLEU-4, and the scores will serve as our optimization target $Y$ for the model train.
The trained ConvNet can predict the commit message from $\{msg_t, msg_g\}$, where \liu{a} higher score means higher relevance of the message to the \texttt{diff}.

\begin{table*}[]
 \centering
\caption{Comparison results with baseline models and different modules within \tool.}
\label{tbl-baseline}
\begin{tabular}{cc|c|c|c|c|c|c}
\toprule 
\multicolumn{2}{c|}{\textbf{Methods}}  &  \liu{\textbf{BLEU-1}} & \liu{\textbf{BLEU-2}} & \liu{\textbf{BLEU-3}} & \textbf{BLEU-4} & \textbf{ROUGE-L} & \textbf{Meteor}   \\ 
\midrule
\multicolumn{1}{c|}{\multirow{5}{*}{\textbf{Baselines}}} &
\multicolumn{1}{c|}{NMT\textsubscript{(Luong)}} 
& 13.12 & 8.01 & 6.11 & 5.23 & 12.73  & 10.37
\\\multicolumn{1}{c|}{} & \multicolumn{1}{c|}{NMT\textsubscript{(Bahdanau)}}    
& 12.78 & 7.66 & 5.72 & 4.81 & 11.95 &  9.87
\\\multicolumn{1}{c|}{} & \multicolumn{1}{c|}{NNGen}   
& 16.91 & 12.01 & 10.03 & 8.04 & 15.20 &  13.68
\\\multicolumn{1}{c|}{} & \multicolumn{1}{c|}{Ptr-Net} 
& 5.80 & 1.72 & 0.73 & 0.45 & 7.61  &  4.98 
\\\multicolumn{1}{c|}{} & \multicolumn{1}{c|}{CODISUM} 
& 7.82 & 3.61 & 2.22 & 1.75 & 9.87   & 8.35  
\\\multicolumn{1}{c|}{} & \multicolumn{1}{c|}{\liu{Commit2Vec}} 
& \liu{12.72} & \liu{7.78} & \liu{6.09} & \liu{5.38} & \liu{13.54} & \liu{10.43}   
\\ 
\midrule
\multicolumn{1}{c|}{\multirow{3}{*}{\textbf{Ours}}}  & \multicolumn{1}{c|}{\tool \textsubscript{Gen}}  
& 15.97 & 10.70 & 8.83 & 7.35 & 14.80 &  11.82
\\ 
\multicolumn{1}{c|}{} & \multicolumn{1}{c|}{\tool \textsubscript{Ret}}   
& 17.74 & 12.65 & 10.55 & 8.52 & 15.93 & 14.35
\\
\multicolumn{1}{c|}{} & \multicolumn{1}{c|}{\tool}   
& \textbf{23.88} &\textbf{15.61}&\textbf{12.17}&\textbf{10.51} & \textbf{22.02} & \textbf{18.51} 
\\
\bottomrule
\end{tabular}
\end{table*}

\section{Experimental Evaluation}\label{sec:expe}
In this section, we conduct experiments to evaluate the effectiveness of \tool and compare it with some state-of-the-art approaches.

% \begin{itemize}%\bfseries
% \item \textbf{RQ1}: What is the performance of \tool comparing with baseline approaches?
% \sql{\item \textbf{RQ2}: What is the impact of the individual module i.e., generation module, retrieval module on the performance of \tool? }
% \item \textbf{RQ3}: How accurate is \subtool under a different number of AST paths?
% \item \textbf{RQ4:} What is the impact of different ranking methods?
% \end{itemize}

\subsection{Setup}
\subsubsection{Experimental Benchmark}
The dataset utilized in previous works~\cite{DBLP:conf/kbse/JiangAM17,DBLP:conf/ijcai/Xu00GT019,DBLP:conf/kbse/LiuXHLXW18} contains no commit ids or complete functions and we can not use directly as ASTs are not available. We crawled 56 popular projects including Neo4j~\cite{neo4j}, Structs~\cite{structs}, Antlr4~\cite{antlr4} from GitHub based on the ``project stars``. The raw messages from this dataset are quite noisy since some commits are empty or contain non-ASCII messages. Furthermore, the merge or rollback commits may contain too many lines, which is not suitable for the generation module. So we filter them out to eliminate unrelated information and remain with 628,887 commits. Additionally, some commits related to project initialization and fundamental functionality updating contain many changes, we remove them \sq{as well}. Specifically, we set the thresholds of chunks as 5 and leave with 438,665 commits. As we need to extract the modified ASTs from java functions so we keep commits with \textit{.java} files and remain 197,968 commits. After removing message length greater than 20 and the same contents of the commits, we keep $\sim$160k samples finally and similar
to Jiang et al. ~\cite{DBLP:conf/kbse/JiangAM17}’s work, randomly select 10\% for testing, 10\% for validation and the remaining for training. For more details about our benchmark, please refer to Section ~\ref{sec:benchmark}.

\subsubsection{Experimental Settings}
For \textit{AST2seq} in the generation module, the max number of paths in \textit{added} and \textit{deleted} ASTs are set to 80 (with more details illustrated in Section 4.3.3). The embedding sizes for subtokens, paths and target sources are defined as 128. The bidirectional LSTM is utilized for encoder layer and LSTM is used for the decoder. All dimensions of the hidden states in the encoder and decoder are fixed to 256. The probability of dropout~\cite{gal2016dropout} is set as 0.4 to avoid overfitting. We set the number of epochs equal to 3,000, along with the batch size as 256 and patience \liu{, a threshold to terminate training for early stopping, as 20.} The learning rate is equal to 0.0001. During testing, we use beam search with beam width as 5 since it has proven useful in sequence prediction with recurrent neural network~\cite{bengio2015scheduled}.
For ConvNet training, we \liu{adopt} a 2-D convolutional layer with the number of kernels defined as
16 and kernel size as $(3, 3)$, followed by a ReLU function and a max-pooling layer with $stride$ size equal to $(2, 2)$. After the max-pooling operation followed by fully connected layers to convert the vector into score values. The optimizer we choose for \textit{AST2seq} and ConvNet is Adam~\cite{kingma2014adam}. 
\liu{We use Tensorflow 1.12~\cite{abadi2016tensorflow} and Pytorch 1.4~\cite{paszke2019pytorch} for our model training.
Hyperparameters such as learning rate, embedding size, encoder and decoder layer numbers, and kernel sizes are tuned with grid search on the validation set~\cite{franceschi2017forward}. The remaining hyperparameters (e.g., beam width and batch size) are configured the same as those in Code2seq~\cite{alon2018code2seq}.} The experiments have been conducted on servers with 36 cores and 4 Nvidia Graphics Tesla P40 and M40. 

\subsubsection{Evaluation Metrics} 
We evaluate our proposed \tool with widely-used automatic metrics such as
\liu{BLEU-N}~\cite{papineni2002bleu}, \liu{ROUGE-L}~\cite{lin2004rouge}, and Meteor
~\cite{banerjee2005meteor}. These metrics have been proved in measuring text similarities between the produced messages and ground truths. 

\liu{BLEU-N computes the n-gram precision of a candidate sequence to the reference, with a penalty for overly short sentences. BLEU-1/2/3/4 correspond to the scores of unigram, 2-grams, 3-grams and 4-grams, respectively}
    \begin{equation}
    BLEU \textnormal{-} N = BP * exp(\sum_{n=1}^{N}w_nlogp_n),
    \end{equation}
    where $N=1,2,3,4$, uniform weights $w_n = 1/N$, and
    \begin{equation}
        BP=\begin{cases}
        1, & \text{if $c>r$}.\\
        e^{1-r/c}, & \text{if $c\leq r$}.
    \end{cases}
    \end{equation}
    where $c$ is the length of the candidate sequence and $r$ is the length of the reference sequence.

\liu{ROUGE-L} provides F-score based on Longest Common Subsequence (LCS). It compares the similarity between two given texts in automatic summarization evaluation.

Meteor modifies the precision and recall computation, replacing them with a weighted F-score based on mapping unigrams and a penalty function for incorrect word order. 
    \begin{equation}
        Meteor = F_{mean}(1-Penalty),
    \end{equation}
    where $F_{mean}$ is computed with unigram precision ($P$) and unigram recall ($R$),
    \begin{equation}
        F_{mean} = \frac{10PR}{R+9P}
    \end{equation}
    and $Penalty$ is levied for fragmented matches as the ratio of matched chunk number to matched unigram number :
    \begin{equation}
        Penalty = 0.5 * (\frac{\# chunks}{\# unigrams\_matched})^3
    \end{equation}

% \sql{Since the metrics, e.g., BLEU-4, ROUGE-L, and METEOR, are computed \yun{by automatic evaluation metrics,}
% % on text similarity, 
% we also conduct a human study to evaluate the semantic similarity \yun{in Section~\ref{sec:human}.}} 

% More details about the semantic similarity evaluation will be presented in Section~\ref{sec:human}.}

\subsection{Comparison Methods}
\sql{We evaluate the proposed \tool against baseline models including the state-of-the-art approaches. We divide them into two groups: 1) Retrieval-based approach: NNGen~\cite{DBLP:conf/kbse/LiuXHLXW18}; and 2) Generation-based approach: NMT~\cite{DBLP:conf/kbse/JiangAM17,DBLP:conf/acl/LoyolaMM17, jiang2019boosting}, Ptr-Net~\cite{ liu2019generating}, CODISUM~\cite{DBLP:conf/ijcai/Xu00GT019} and \liu{Commit2Vec~\cite{commit2vec}}. For the implementation, we reproduce NNGen~\cite{DBLP:conf/kbse/LiuXHLXW18} by using the same algorithm and settings according to the original paper.} \liu{The source code of NMT
% ~\footnote{\url{https://github.com/tensorflow/nmt}}
, Ptr-Net~
% \footnote{\url{https://zenodo.org/record/2542706\#.XxPkmvgzZTa}}\cite{liu2019generating} 
and CODISUM
% ~\footnote{\url{https://github.com/SoftWiser-group/CoDiSum}}\cite{DBLP:conf/ijcai/Xu00GT019}
is available online \cite{nmt_code, Ptr-Net_code, CODISUM_code} and we utilize the default settings described in the corresponding papers. For Commit2Vec, we try our best to replicate the code for commit message generation according to the paper and make the replication publicly available
\cite{Commit_code}
% \footnote{\href{https://github.com/shangqing-liu/ATOM}{{https://github.com/shangqing-liu/ATOM}}}.
}
The details of these approaches are illustrated below.

% \liu{The source code of NMT~\footnote{\url{https://github.com/tensorflow/nmt}} \cite{DBLP:conf/kbse/JiangAM17,DBLP:conf/acl/LoyolaMM17, jiang2019boosting}, Ptr-Net~\footnote{\url{https://zenodo.org/record/2542706\#.XxPkmvgzZTa}}\cite{liu2019generating} and CODISUM~\footnote{\url{https://github.com/SoftWiser-group/CoDiSum}}\cite{DBLP:conf/ijcai/Xu00GT019} is available online and we utilize the default settings from the corresponding papers. For Commit2Vec, since it utilized code2vec~\cite{alon2019code2vec} as a building block to represent code changes for security commit classification. We also adapt Commit2Vec for message generation by the released code2vec~\footnote{https://github.com/tech-srl/code2vec}.} The details of these approaches are illustrated below.

\begin{itemize}
    \item NNGen~\cite{DBLP:conf/kbse/LiuXHLXW18}. NNGen is a retrieval-based approach which retrieves the most similar top-\textit{k} \texttt{diff}s from the training dataset based on a bag-of-words~\cite{christopher2008introduction} model and 
    prioritizes the \texttt{diff} candidates in terms of BLEU-4 scores. NNGen regards the message of the \texttt{diff} with the highest BLEU-4 score as the result.
	\item NMT~\cite{DBLP:conf/kbse/JiangAM17,DBLP:conf/acl/LoyolaMM17, van2019generating}. NMT adopts attention-based RNN encoder-decoder models, described in Section~\ref{sec:en-de} to generate commit messages for \texttt{diff}s. Jiang et al.~\cite{DBLP:conf/kbse/JiangAM17} uses Bahdanau attention \cite{bahdanau2014neural} to produce messages. Another approach Commitgen proposed by Loyola et al.~\cite{DBLP:conf/acl/LoyolaMM17} leverages Luong \cite{luong2015effective} attention instead of Bahdanau for commit message generation. We compare both attention mechanisms denoted as NMT\textsubscript{(Luong)} and NMT\textsubscript{(Bahdanau)}. 
% 	\sql{Furthermore, the latest work~\cite{jiang2019boosting} proposed to utilize the results produced by ChangeScribe~\cite{DBLP:conf/scam/Cortes-CoyVAP14,DBLP:conf/icse/VasquezCAP15} to replace original commit messages for model training to boost the generation performance. We also compared with this approach denoted as NMT\textsubscript{(Boost)}.}
	\item Ptr-Net~\cite{DBLP:conf/acl/SeeLM17, liu2019generating}. Ptr-Net (an abbreviation of Pointer network) is a typical text summarization approach, which can copy the Out-of-Vocabulary (OOV) words such as variable and method names from source code to the generated messages. Ptr-Net has proven effective in generating rational commit messages for code changes by Liu et al. \cite{liu2019generating}.
	\item CODISUM~\cite{DBLP:conf/ijcai/Xu00GT019}. CODISUM is the state-of-the-art approach which \sq{employs the normalized code changes in which the identifiers are unified with corresponding placeholders for learning the representations of code changes} as well as combining pointer network~\cite{DBLP:conf/acl/SeeLM17} to mitigate the OOV issue.
	\item \liu{Commit2Vec}\cite{commit2vec}. \liu{Commit2Vec feeds the \textit{added} and \textit{deleted} AST paths to a fully-connected layer to encode code changes for
    classifying security-related commits. Although Commit2Vec is targeted at binary classification, the encoding mode of code changes can be adopted for various downstream tasks including commit message generation, so we also consider Commit2Vec as one baseline. Since the source code is not publicly available, we tried our best to reproduce the model according to the paper.''}
% 	\liu{Commit2Vec is inspired by code2vec~\cite{alon2019code2vec} and feed the \textit{added} and \textit{deleted} paths to the fully-connected layer to compute the combined context vectors as the commit representation for downstream tasks i.g. security commit classification. We reproduce this approach of code change representation and use it for commit message generation.}
\end{itemize}

\subsection{Experimental Results}
We present the experimental results and analysis through the following \liu{research} questions.
\subsubsection{What is the performance of \tool comparing with baseline approaches?}
\sql{
Table~\ref{tbl-baseline} shows the results of our approach against the baselines. We can find that \tool outperforms the baselines by a significant margin. 
% Inspection on the performance of baselines, NMT\textsubscript{(Boost)} has a higher performance as compared to basic NMT models, i.e., NMT\textsubscript{(Luong)}, NMT\textsubscript{(Bahdanau)}. However, by a deep analysis of NMT\textsubscript{(Boost)}, we find it suffers from a serious flaw. NMT\textsubscript{(Boost)} tailored the messages for the NMT input by utilizing ChangeScribe~\cite{DBLP:conf/scam/Cortes-CoyVAP14,DBLP:conf/icse/VasquezCAP15} for the higher performance. However, this operation modifies the original messages to some common templates with less semantics, e.g., ``This commit renames some files.'' ``Changes to a file.'' and use these modified messages for model training and evaluation. It can be considered as a trick for the original data and the generated messages can not represented the originals.
The improved models such as Ptr-Net and CODISUM, which claimed to capture code semantics, have lower performances. In essence, they treat code as a flat sequence of tokens, failing to capture the semantics behind the code. 
\liu{Commit2Vec, which also adopts AST paths to represent code changes, presents lower performance than \tool in terms of all the evaluation metrics. The lower performance may be attributed to that Commit2Vec utilizes a fully-connected layer to represent code changes and could fail to capture the sequential information in the \textit{added}/\textit{deleted} AST paths. In our proposed \subtool,  bi-directional  LSTMs are involved to incorporate the sequential information of the \textit{added}/\textit{deleted} AST paths for better representing code changes.}
The retrieval-based approach NNGen has a higher performance than pure generation approaches, demonstrating the effectiveness of the retrieval-based method on message generation tasks. Finally, \tool improves all the baseline approaches by 30.72\%, 44.89\%, and 35.26\% in terms of BLEU-4, ROUGE-L, and Meteor respectively. This can be attributed to that \tool can effectively integrate the advantages of generation and retrieval modules.
}

\begin{tcolorbox}[breakable,width=\linewidth,boxrule=0pt,top=1pt, bottom=1pt, left=1pt,right=1pt, colback=gray!20,colframe=gray!20]
\textbf{Answer to RQ1:} In summary, \tool improves the baseline approaches by 30.72\%, 44.89\%, 35.26\% in terms of BLEU-4, ROUGE-L, and Meteor respectively.
\end{tcolorbox}

\subsubsection{What is the impact of \sql{individual modules} on the performance of \tool?}
\sql{We also perform experiments to evaluate the impact of individual generation module and retrieval module on the generated commit messages, with the results shown in Table~\ref{tbl-baseline}. No ranking model is included in the two variants of \tool.}
We denote the results produced only by the retrieval module as ATOM\textsubscript{Ret}, which uses TF-IDF to retrieve the most similar commit message (see Section~\ref{sec:ret}), and generation module as ATOM\textsubscript{Gen}, which only uses \textit{AST2seq} for commit message generation (see Section~\ref{sec:gen}). We find that the performance of ATOM\textsubscript{Ret} is slightly higher than ATOM\textsubscript{Gen}, but the overall performance is still lower than the combined model \tool. The gains achieved by our hybrid ranking module range from 12.76\% to 56.58\% in terms of BLEU-4, ROUGE-L and Meteor. Hence, \tool incorporates the retrieval results into the generated results by \liu{the} hybrid ranking module will further boost the performance. In addition, ATOM\textsubscript{Gen} achieves the best performance among all the generation approaches, i.e., NMT, Ptr-Net, and CODISUM, which proves that utilizing AST to learn code semantics is more powerful than simple sequential models. Finally, compared with the retrieval-based approach NNGen, ATOM\textsubscript{Ret} has slightly better performance, since we retrieve the most similar commit message based on the weight of tokens. Hence, some important tokens with low frequency will be considered, which is superior to NNGen.

As \tool outputs the commit messages produced by either generation or retrieval module, we also analyze the proportions of the messages from each module, with statistics shown in Table~\ref{tbl-ret-gen}. In a total of 14,674 testing samples, 8,168 of the results are from the retrieval module, accounting for 55.66\% of the entire testing corpus and the remaining 6,506 are from the generation module (44.34\%). Based on the statistics, we can conclude that both retrieval and generation modules are helpful for accurate commit message generation, and they are \liu{complementary} to each other.

\begin{tcolorbox}[breakable,width=\linewidth,boxrule=0pt,top=1pt, bottom=1pt, left=1pt,right=1pt, colback=gray!20,colframe=gray!20]
\textbf{Answer to RQ2:} In summary, \tool incorporates the retrieval results into generation module to boost the final performance, and the improvement range from 12.76\% to 56.58\% in BLEU-4. Furthermore, among all the generation approaches, our proposed \subtool can learn more semantics in the commits to produce high-quality messages.
\end{tcolorbox}

% \begin{table*}[t]
%   \centering
% %   \addtolength{\tabcolsep}{-3.5pt}
% 	\caption{Impact of different modules}
% 	\label{tbl-module}
	
% % 	\scalebox{1.0}{
% 	\begin{threeparttable}
% 	\begin{tabular}{c| l| c c c c c}
% 	  \toprule 
% 	  \multicolumn{2}{c|}{\textbf{Model}} & \textbf{BLEU-4} & \textbf{ROUGE-1}  & \textbf{ROUGE-2} & \textbf{ROUGE-L} & \textbf{Meteor}\\
% 	  \midrule
% 	  \multirow{2}{*}{\begin{tabular}{@{}c@{}}\textbf{ATOM with each} \\ \textbf{module }\end{tabular}} 
% 	  & \tool \textsubscript{Gen} & 7.35 & 0.17 & 0.06 & 0.15 & 0.12  \\
% 	  & \tool \textsubscript{Ret} & 8.52 & 0.17 & 0.08 & 0.16 & 0.14  \\
% 	  \midrule
% 	  \multicolumn{2}{c|}{ATOM} & \textbf{10.51} & \textbf{0.22} & \textbf{0.10} & \textbf{0.22} & \textbf{0.19} \\
%       \bottomrule
%       \end{tabular}
%       \label{tbl-module}
% 	\end{threeparttable}
% % 	}
% \end{table*}

\subsubsection{How accurate is \subtool under a different number of paths?}\label{path-num}

\begin{figure}[t]
     \centering
     \includegraphics[width=0.5\textwidth]{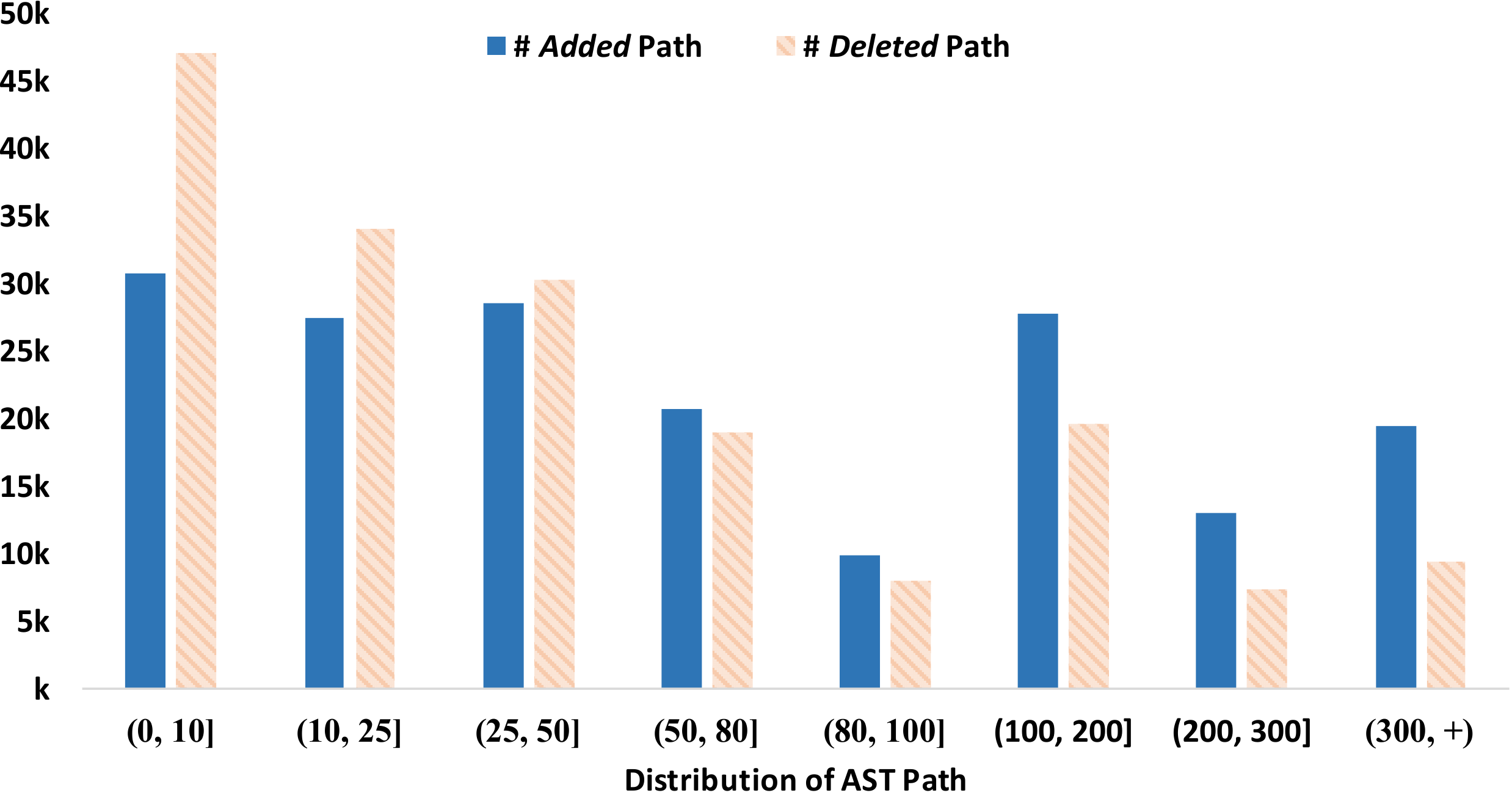}
     \caption{The distribution of AST paths shown in the dataset. Each bar represents the number of commits that has the number of AST paths in a specific interval. For example, the leftmost blue bar represents almost 30,000 commits in our dataset have less than 10 \textit{added} AST paths by our preprocess.} 
     \label{fig:chart}
\end{figure}

Our generation module \subtool encodes AST paths based on \texttt{diff}s to represent code changes, however, the number of paths vary depending on the length of \texttt{diff}s. In this paper, we set the max number of paths to 80 for the \textit{added} and \textit{deleted} ASTs during training respectively. From Fig.~\ref{fig:chart}, we can see that nearly 80\% of commits have fewer than 80 AST paths in our dataset. \sql{In this RQ, we analyze the impact of different numbers of AST paths on the model performance. Specifically, we truncate the ASTs with longer paths to be the experimental number of ASTs. For example, to examine the results when taking the number of AST paths as 30, we randomly select 30 paths for the ASTs with real paths larger than 30. The results are illustrated in Table~\ref{tbl-paths}.} As can be seen, the optimal value of the path number in our experiment is 80 and BLEU-4, ROUGE(1,2,L) and Meteor achieves 7.35, 16.69, 6.23, 14.80 and 11.82 respectively. Furthermore, few path numbers tend to show worse results, e.g., when the path is set as 30, the performance decreases dramatically to 4.27. It can be attributed to fewer paths have limited capability in representing code changes. Increasing paths to over 100 \liu{do} not result in continuously improved performance and the scores show \liu{a} slight decrease when the paths augmented from 200 to 300. In addition, large numbers of paths will be a heavy burden for model training. Hence, we can conclude that 80 is an optimal value to represent \texttt{diff}s.

\begin{tcolorbox}[breakable,width=\linewidth,boxrule=0pt,top=1pt, bottom=1pt, left=1pt,right=1pt, colback=gray!20,colframe=gray!20]
\textbf{Answer to RQ3:} Overall, the optimal value of AST paths for effectively representing
\texttt{diff}s is 80. Adding fewer or more paths cannot contribute much to the performance.
\end{tcolorbox}

\begin{table}[t]
  \centering
	\caption{Percentage of final results prioritized from retrieved and generated messages.}
	\label{tbl-ret-gen}
	\scalebox{1.0}{\begin{threeparttable}
	\begin{tabular}{c | c c}
	  \toprule 
    \textbf{Modules} & \textbf{Number} & \textbf{Percentage (\%)} \\
    \midrule
     \textbf{Retrieval} & 8,168 &55.66\%\\
     \textbf{Generation} & 6,506 & 44.34\%\\
    \bottomrule
	\end{tabular}
    \end{threeparttable}
    }
\end{table}

% \begin{table}[t]
%   \centering
% %   \addtolength{\tabcolsep}{-3.5pt}
% 	\caption{\sql{The performance of path set on \textit{AST2seq}. The left column represents the path number for \textit{added} and \textit{deleted} paths separately.}}
% 	\label{tbl-paths}
% 	\scalebox{0.87}{\begin{threeparttable}
% 	\begin{tabular}{c | c c c c c c }
% 	  \toprule 
%     \textbf{\# Path} %\textit{AST2seq
%     & \textbf{BLEU-4} & \textbf{ROUGE-1} &
%      \textbf{ROUGE-2} & \textbf{ROUGE-L} &  \textbf{Meteor}\\
%     \midrule
%      30  & 4.27 & 15.55 & 4.50  & 13.66 & 10.22 \\
%      50  & 5.96 & 15.41 & 5.26  & 13.92 & 10.89 \\
%      \textbf{80} &\textbf{7.35} & \textbf{16.69}  & \textbf{6.23} & \textbf{14.80} & \textbf{11.82} \\
%      100 & 7.09 & 16.40 & 5.98  & 14.34 & 11.42 \\
%      200 & 6.07 & 15.47 & 5.30  & 14.01 & 11.01 \\
%      300 & 6.04 & 15.38 & 5.27  & 13.95 & 10.99 \\
%     %  \midrule
%     %  60  & 6.80 & 15.90 & 5.73  & 14.24 & 11.15     \\
%     %  70  & 6.95 & 16.01 & 5.82  & 14.32 & 11.21     \\
%     %  90  & 7.21 & 16.58 & 6.07  & 14.56 & 11.56     \\
%     %  \midrule
%     \bottomrule
% 	\end{tabular}
%     \end{threeparttable}
%     }
% \end{table}

\begin{table}[t]
  \centering
%   \addtolength{\tabcolsep}{-3.5pt}
	\caption{\sql{The performance of path set on \textit{AST2seq}. The left column represents the path number for \textit{added} and \textit{deleted} paths separately.}}
	\label{tbl-paths}
	\scalebox{0.8}{\begin{threeparttable}
	\begin{tabular}{c | c c c c c c c}
	  \toprule 
    \textbf{\# Path} %\textit{AST2seq
    & \liu{\textbf{BLEU-1}} & \liu{\textbf{BLEU-2}} & \liu{\textbf{BLEU-3}} &
     \textbf{BLEU-4} & \textbf{ROUGE-L} &  \textbf{Meteor}\\
    \midrule
     30  & 11.53 & 6.57 & 4.80 & 4.27 & 13.66 & 10.22 \\
     50  & 13.67 & 8.70 & 6.83 & 5.96 & 13.92 & 10.89 \\
     \textbf{80} & \textbf{15.97} & \textbf{10.70} & \textbf{8.83} &\textbf{7.35} & \textbf{14.80} & \textbf{11.82} \\
     100 & 15.11 & 10.01 & 8.19 & 7.09 & 14.34 & 11.42 \\
     200 & 13.89 & 8.83  & 6.88 & 6.07 & 14.01 & 11.01 \\
     300 & 13.72 & 8.77  & 6.85 & 6.04 & 13.95 & 10.99 \\
    %  \midrule
    %  60  & 6.80 & 15.90 & 5.73  & 14.24 & 11.15     \\
    %  70  & 6.95 & 16.01 & 5.82  & 14.32 & 11.21     \\
    %  90  & 7.21 & 16.58 & 6.07  & 14.56 & 11.56     \\
    %  \midrule
    \bottomrule
	\end{tabular}
    \end{threeparttable}
    }
\end{table}

% \begin{table}[t]
%   \centering
% %   \addtolength{\tabcolsep}{-3.5pt}
% 	\caption{The performance of different ranking methods }
% 	\label{tbl-classifier}
% 	\scalebox{0.87}{\begin{threeparttable}
% 	\begin{tabular}{c | c c c c c c c}
% 	  \toprule 
%     \textbf{Methods} & \textbf{BLEU-4} & \textbf{ROUGE-1} &
%      \textbf{ROUGE-2} & \textbf{ROUGE-L} &  \textbf{Meteor}\\
%     \midrule
%      XGBoost & 8.93 & 17.17 & 7.78 & 15.48 & 13.82 \\
%      SVR  & 8.73 & 16.88 & 7.61 & 15.22 & 13.46\\
%      GRU  & 9.34 & 17.37 & 8.17 & 15.72 & 14.10 \\
%      LSTM & 9.32 & 17.51 & 8.22 & 15.85 & 14.18 \\
%      LSTM+Att & 9.33 & 17.49 & 8.18 & 15.82 & 14.16 \\
%      \midrule
%      ConvNet &\textbf{10.51} & \textbf{24.33}  & \textbf{9.55} & \textbf{22.02} & \textbf{18.51} \\
%     \bottomrule
% 	\end{tabular}
%     \end{threeparttable}
%     }
% \end{table}

\begin{table}[t]
  \centering
%   \addtolength{\tabcolsep}{-3.5pt}
	\caption{The performance of different ranking methods. }
	\label{tbl-classifier}
	\scalebox{0.8}{\begin{threeparttable}
	\begin{tabular}{c | c c c c c c c c}
	  \toprule 
    \textbf{Methods} & \liu{\textbf{BLEU-1}} & \liu{\textbf{BLEU-2}}  &
     \liu{\textbf{BLEU-3}} & \textbf{BLEU-4} & \textbf{ROUGE-L} &  \textbf{Meteor}\\
    \midrule
     XGBoost   &17.61 &12.21 &10.01 & 8.93  & 15.48 & 13.82 \\
     SVR       &16.99 &11.83 &9.74  & 8.73  & 15.22 & 13.46\\
     GRU       &17.64 &12.48 &10.37 & 9.34  & 15.72 & 14.10 \\
     LSTM      &17.70 &12.49 &10.36 & 9.32  & 15.85 & 14.18 \\
     LSTM+Att  &17.74 &12.51 &10.37 & 9.33  & 15.82 & 14.16 \\
     \midrule
     ConvNet &\textbf{23.88} &\textbf{15.61}&\textbf{12.17}&\textbf{10.51} & \textbf{22.02} & \textbf{18.51} \\
    \bottomrule
	\end{tabular}
    \end{threeparttable}
    }
\end{table}

\subsubsection{What is the impact of different ranking methods?} \label{sec:ranking}
\tool designs a ConvNet to incorporate the output of retrieval module into generation module \subtool to get better results. However, the hybrid ranking module can be regarded as a regression problem, and be solved with other alternatives. In this section, we evaluate the performance of different methods for the ranking module. We choose XGBoost~\cite{chen2016xgboost}, Support Vector Regression (SVR)~\cite{drucker1997support}, GRU \cite{chung2014empirical} and LSTM \cite{hochreiter1997long} with or without Attention Mechanism as the baselines for ranking. We compute tf-idf scores for the tokens in messages and \texttt{diff}s as features for training XGBoost and SVR. For the other baselines, we concatenate the hidden states of messages and \texttt{diff}s as the feature representations and then fed into a fully-connected layer for predicting the relevance score. 
\sql{The training loss functions are similar to the definition in ConvNet and all the hyperparameters are well-tuned by grid search~\cite{franceschi2017forward}.}
The comparison results are shown in Table~\ref{tbl-classifier}. We can find that deep learning methods i.e., GRU, LSTM, LSTM+Att, outperform machine learning methods i.e., XGBoost, SVR. Specifically, XGBoost presents a better performance as compared to SVR as it combines a set of classification and regression trees (CART)~\cite{lewis2000introduction} to gradually reduce prediction errors by each iteration. 
\sql{
The superior performance of ConvNet is because ConvNet adopts the similarity matching matrix (Section~\ref{sec:matrix}) to directly capture the relevance between \texttt{diff}s and candidate messages, for message ranking instead of concatenating their respective representations.
}
\begin{tcolorbox}[breakable,width=\linewidth,boxrule=0pt,top=1pt, bottom=1pt, left=1pt,right=1pt, colback=gray!20,colframe=gray!20]
\textbf{Answer to RQ4:} 
For predicting relevance between candidate messages produced by generation module and retrieval module, ConvNet is superior to traditional machine learning models, {e.g.}, XGBoost and SVR, and sequential deep learning models, {e.g.}, GRU and LSTM. 
\end{tcolorbox}

\subsection{Human Study}\label{sec:human}
We conduct a human evaluation to evaluate \tool with the best retrieval model NNGen~\cite{DBLP:conf/kbse/LiuXHLXW18} and the best generation model NMT\textsubscript{(Luong)}~\cite{DBLP:conf/acl/LoyolaMM17}. 
% \sql{Here, we discard NMT\textsubscript{(Boost)} even it has a higher BLEU-4 than NMT\textsubscript{(Luong)}~\cite{DBLP:conf/acl/LoyolaMM17}, since it tailored original commit messages to some meaningless messages, which is not a corrected practise for this task. }
We invite 4 PhD students and 2 master students from the department of computer science to participate in our survey. None of the participants \liu{is} co-authors of this paper and they all have software development experience in Java programming language (raging 1 $\sim$ 5 years).

\subsubsection{Survey Design}
\begin{figure}[t]
     \centering
     \includegraphics[width=0.5\textwidth]{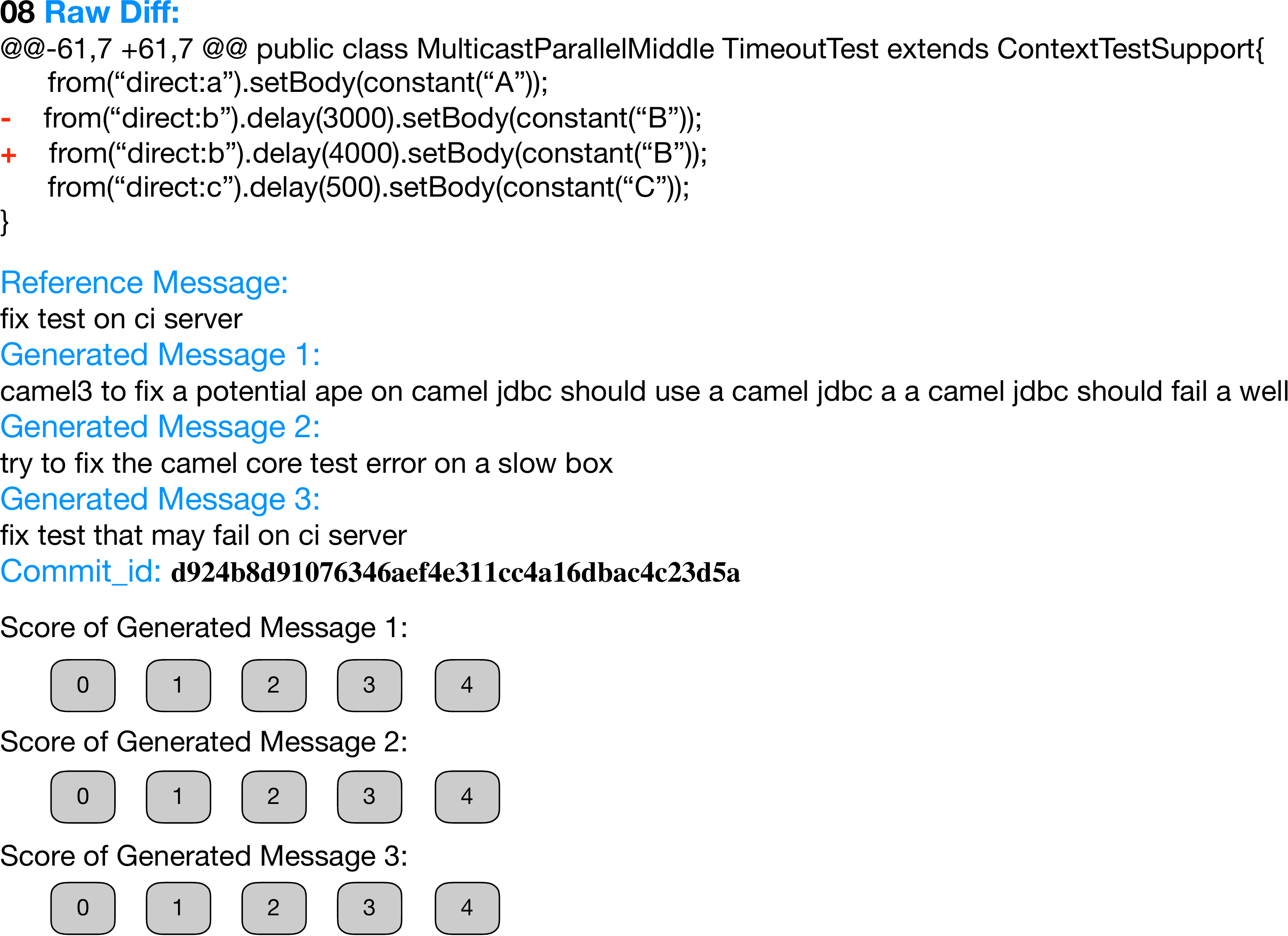}
     \caption{A case of the questionnaire, provided with RAW Diff, followed by Reference Message and Generated Messages to score. We also provide commit id in case of participants to search on the Internet.} 
     \label{fig:questionare}
\end{figure}

We randomly selected 100 commits from the test dataset for each participant to read and assess. In our questionnaire, each question first presents the code changes of one commit, i.e., its \texttt{diff}, its reference message, and messages produced by NNGen, NMT\textsubscript{(Luong)}, and \tool respectively. Each participant is asked to give three quality scores between 0 to 4 to indicate the semantic similarities between the reference message and the three generated messages. Lower scores mean the generated messages are less identical to the reference messages. Fig.~\ref{fig:questionare} shows one question in our survey. Participants are told the first message is the reference message, but the others are not aware of which message is generated by which approach and the three messages are randomly ordered. They are asked to score each generated message separately. Furthermore, we provide the commit id to help participants to search related information through the Internet.

\begin{table*}[t]
  \centering
%   \addtolength{\tabcolsep}{-3.5pt}
	\caption{\sql{The score distribution of the generated commit messages by NNGen, NMT\textsubscript{(Luong)}, and \tool. The standard deviation is illustrated beside the average score.}}
	\label{tbl-human}
	\begin{tabular}{c | c c c c c| c c c |c |c | c}
	 \toprule 
    \textbf{Methods} & 0 & 1 & 2 & 3 & 4 & $\geq$ 3 & $\geq$ 2 & $\leq$ 1  & Avg. (STD) & BLEU-4 & PCC\\  
    \midrule
    NNGen &22  & 28 & 27 & 15 & 8  & 23  & 50  & 50  & 1.59 (0.68) & 8.81 &  0.17
  \\
    NMT\textsubscript{(Luong)}   &32  & 26 & 26 & 10 & 6  & 16  & 42  & 58  & 1.32 (0.83) & 6.17 & 0.06
  \\
    ATOM  &18  & 25 & 31 & 16 & 10 & 26  & 57  & 43  & 1.75 (0.59) & 10.23 &  0.21\\
    \bottomrule
	\end{tabular}
\end{table*}

\subsubsection{Survey Results}
Each code change and commit message pair is evaluated by 6 participants. Our scoring criterion is listed at the beginning of each questionnaire to guide participants, which follows Liu et al.'s work~\cite{DBLP:conf/kbse/LiuXHLXW18}, e.g., \sq{score 0 means two messages have no shared tokens and score 1 denotes they have some shared tokens, but without semantic similarity.} Score 2 can have some similar information but lacking important parts and score 3, 4 denotes two messages are very similar in semantics or even identical. We finally obtain 600 pairs of scores from our human evaluation. Each pair contains corresponding scores for the messages generated by NMT\textsubscript{(Luong)}, NNGen, and ATOM, respectively.
\sql{
Table~\ref{tbl-human} shows the score distribution of the generated commit messages based on the three methods. We can find that \tool receives the best score and improve the average scores of NMT\textsubscript{(Luong)} (1.32) and NNGen (1.59) to 1.75. Furthermore, our approach can generate more high-quality messages (scores $\geq$ 3) than NNGen and NMT\textsubscript{(Luong)}. By comparing with the quantitative evaluation results of the 100 sampled commits, i.e., the BLEU-4 scores listed in the last column of Table~\ref{tbl-human}, we can observe that they are consistent with the human ranking results. We then employ Pearson Correlation Coefficient (PCC)~\cite{pearson1895notes} to compute the correlations between the manual annotations and corresponding BLEU-4 scores for the 100 commits. The results also show that the messages generated by ATOM receive the most consistent scores between human study and the automated evaluation, with PCC score at 0.21. For the messages generated by NNGen and NMT\textsubscript{(Luong)}, the PCC scores are relatively lower, at 0.17 and 0.06, respectively.
}

\sql{We also conduct inter-rater agreement analysis~\cite{hallgren2012computing} on the manual annotations to observe the consistency among participants. We find that the agreement rates for the studied three approaches NNGen, NMT\textsubscript{(Luong)}, and \tool are 51.78\%, 45.32\%, and 62.60\%, respectively. Considering each commit is annotated by six participants and the difficulty of the task, the agreement rates are reasonable and acceptable~\cite{DBLP:conf/kbse/LiuXHLXW18}. The result also implies that the generated messages by \tool present the highest quality among all the generated messages.}

\begin{tcolorbox}[breakable,width=\linewidth,boxrule=0pt,top=1pt, bottom=1pt, left=1pt,right=1pt, colback=gray!20,colframe=gray!20]
Overall, by our human study, we can conclude that \tool can produce more semantically related results with the ground-truths.
\end{tcolorbox}

\subsection{\liu{Examples}}\label{sec:case}
\sql{\liu {We show some examples} to analyze the strengths and weaknesses of \tool. Two examples of the generated messages by \tool, NNGen, NMT\textsubscript{(Luong)}, and the ground truth are illustrated in Table~\ref{tbl-example}. From Example 1, we can find that both messages generated by NNGen and NMT\textsubscript{(Luong)} fail to describe the code change. For NNGen, since it directly recommends the message of the \texttt{diff} from the training set, it may fail when no relevant \texttt{diff}s appear in the training set. The generated message by NNGen contains words such as ``camel3'' and ``npe'' which are obviously unrelated to the \texttt{diff}. For NMT\textsubscript{(Luong)}, it uses a sequence of code tokens as input, which may not accurately capture the semantics of a code change. As shown in Example 1,  NMT\textsubscript{(Luong)} does not recognize that the \texttt{diff} is used to fix test. In contract, ATOM utilizes ASTs to capture the semantics of the \texttt{diff} and can generate a more accurate commit message.}

\sql{As shown in Example 2 of Table~\ref{tbl-example}, all the generated messages fail to detail that the code change is related to ``jmstype header''. This may be because the textual information, e.g., logs, in code changes is not well exploited and attended. In future, we will adopt text mining techniques such as part-of-speech analysis to fully capture the semantics in textual information of \texttt{diff}s.}

\begin{table*}[t]
% \scriptsize{ \addtolength{\tabcolsep}{-4pt}
\centering
\caption{\sql{Examples of generated commit messages.}}
\label{tbl-example}
\begin{adjustbox}{width=\textwidth}
\begin{tabular}{c|c|c}
\hline
    Example & Example 1 & Example 2 \\
    \hline
    Commit\_id & d924b8d91076346aef4e311cc4a16dbac4c23d5a & 
    47a7eabe91fc8e8d26518bd092e62cdc7570d9af\\
    \hline
    \texttt{diff} & 
    \begin{lstlisting}[basicstyle=\small]
    @@ -61,7 +61,7 @@ 
    public class MulticastParallelMiddleTimeoutTest 
    extends ContextTestSupport {
      from("direct:a").setBody(constant("A"));
    - from("direct:b").delay(3000)
      .setBody(constant("B"));
    + from("direct:b").delay(4000)
      .setBody(constant("B"));
      from("direct:c").delay(500)
      .setBody(constant("C"));
    }
    \end{lstlisting}   &
    \begin{lstlisting}[basicstyle=\small]
    @@ -183,7 +183,7 @@ 
    public class JmsBinding {
    try {
        map.put("JMSType", jmsMessage.getJMSType());
    } catch (JMSException e) {
    -     LOG.trace("Cannot read JMSReplyTo header. 
          Will ignore this exception.", e);
    +     LOG.trace("Cannot read JMSType header. 
          Will ignore this exception.", e);
    }
    \end{lstlisting}
    
    \\
    \hline
    Ground-Truth  & fix test on ci server & oracleaq do not support jmstype header\\
    \hline
    NNGen & camel3 to fix a potential npe on camel jdbc & oracleaq do not work\\
    \hline
    NMT\textsubscript{(Luong)} & try to fix the camel core test error on a slow box & ensure stack trace be in trace log of exception\\
    \hline
    \textbf{ATOM}  & \textbf{fix test that may fail on ci server} &\textbf{oracleaq do not support} \\
\hline
\end{tabular}
\end{adjustbox}
\end{table*}

\section{Discussion}\label{sec:discuss}
In this section, we describe the strengths of \textit{AST2seq} as compared to NMT approaches, then provide more details about our benchmark compared with Jiang's ~\cite{DBLP:conf/kbse/JiangAM17} , and present the difference with Code2seq and Commit2Vec. \sq{Then we give a discussion about OOV issue in \tool} and finally discuss the limitations of \tool.

\subsection{Strengths of \textit{AST2seq}}
Previous studies, e.g., NMT~\cite{DBLP:conf/kbse/JiangAM17,DBLP:conf/acl/LoyolaMM17}, Ptr-Net~\cite{liu2019generating} treated \texttt{diff} as a flat sequence of tokens, which ignored code semantic information. To address this limitation, CODISUM~\cite{DBLP:conf/ijcai/Xu00GT019} extracted code structure and code semantics based on identifying all the class/method/variable names and segmenting with the corresponding placeholders. By this way, they achieved BLEU-4 of 2.19 on Jiang's~\cite{DBLP:conf/kbse/JiangAM17} dataset. Although they claimed that they achieved the highest BLEU-4 over NMT methods on Jiang's dataset, the performance is still far away from satisfaction, which encourages us to do further exploration. 

Many methods with the same functionality by a different implementation tend to have different surface forms, which is particularly common in the ``For`` and ``While`` statements. However, NMT-based approaches essentially treat \texttt{diff}s as a sequence of tokens, which hinders from capturing the semantics as the diverse expression format. However, AST is a high abstraction of code snippet and it transfers methods from plain text to tree structure. In many cases, methods with the same functionality share similar AST structures. Therefore, encoding AST to learn code semantics can seem as a refinement of original source codes and the recurring patterns might be easier to capture. 

In addition, to easily capture semantics among \texttt{diff}s to represent code changes, another advantage for \textit{AST2seq} is the ability to handle longer code changes.  In sequence-based approaches~\cite{DBLP:conf/kbse/JiangAM17,DBLP:conf/acl/LoyolaMM17, liu2019generating, DBLP:conf/ijcai/Xu00GT019}, they need to set maximum sequence, e.g., 100 tokens in total~\cite{DBLP:conf/kbse/JiangAM17} for effective learning, which will lead to filter out a commit with too many chunks. Hence the sequence-based models cannot translate a commit with long sequences. However, \textit{AST2seq} can effectively address this limitation. We extract paths between leaf nodes and combine them to represent code changes instead of treating \texttt{diff}s as a flat sequence. The number of sampled paths in \textit{added} and \textit{deleted} ASTs is set to 80 separately and the larger will be truncated. By this way, \textit{AST2seq} is able to handle longer \texttt{diff}s and the results about the performance among different \texttt{diff} length are shown in Table~\ref{tbl-difflen}, where the left column is the \texttt{diff} lines rather than token length. The BLEU-4  within 10 lines \texttt{diff}s is 11.37 and it takes up 13.60\% in the whole testset. When diff lines increase to 100+, the performance only decreases by 2.13, 1.92, and 2.65 in terms of BLEU-4, ROUGE-L, and Meteor.
Moreover, the performance with lines within 50-100 is better than the lines within 10-30 and 30-50. Therefore, the performance will not decrease dramatically along with the increased \texttt{diff} lines. Hence, \textit{AST2seq} uses ASTs to encode code changes addressing the limitation of sequence length.
\begin{tcolorbox}[breakable,width=\linewidth,boxrule=0pt,top=1pt, bottom=1pt, left=1pt,right=1pt, colback=gray!20,colframe=gray!20]
\textbf{} 
%\noindent 
To sum up, \textit{AST2seq} utilizes AST into the encoder to learn the semantics behind the code changes and can handle longer \texttt{diff}s, which is superior to the existing approaches.
\end{tcolorbox}

% \begin{table*}[h]
%   \centering
% 	\caption{Results of \texttt{diff}s with different lines rather than tokens. For example the upper left \textbf{1-10} in the \texttt{diff} Lines column represents the commits with at most 10 lines of \texttt{diff}s.}
% 	\label{tbl-difflen}
% 	\scalebox{0.87}{\begin{threeparttable}
% 	\begin{tabular}{c | c c c c c c c}
% 	  \toprule 
%     \textbf{\texttt{diff} lines} & \textbf{BLEU-4} & \textbf{ROUGE-1} &
%      \textbf{ROUGE-2} & \textbf{ROUGE-L} &  \textbf{Meteor} & \textbf{Number} & \textbf{Ratio}\\
%     \midrule
%     \textbf{1-10} & \textbf{11.37} & \textbf{25.88} & \textbf{10.84} & \textbf{23.48} & \textbf{20.41} & 2920 & 19.90\%  \\
%     \textbf{10-30} & 9.93 & 23.00 & 9.01 & 20.77 & 17.71 & 3435 & 23.41\%  \\
%     \textbf{30-50} & 9.33 & 24.15 & 9.29 & 21.83 & 18.63 & 2652 & 18.07\%   \\
%     \textbf{50-100} & 10.86 & 24.94 & 9.97 & 22.56 & 19.19 & \textbf{3670} & \textbf{25.01}\%      \\
%     \textbf{100+} & 9.37 & 23.62 & 8.72 & 21.27 & 17.77 & 1996 & 13.60\%      \\
%     \bottomrule
% 	\end{tabular}
%     \end{threeparttable}
%     }
% \end{table*}

\begin{table*}[h]
  \centering
	\caption{Results of \texttt{diff}s with different lines rather than tokens. For example the upper left \textbf{1-10} in the \texttt{diff} Lines column represents the commits with at most 10 lines of \texttt{diff}s.}
	\label{tbl-difflen}
	\scalebox{0.87}{\begin{threeparttable}
	\begin{tabular}{c | c c c c c c c c}
	  \toprule 
    \textbf{\texttt{diff} lines} & \liu{\textbf{BLEU-1}} & \liu{\textbf{BLEU-2}} &
     \liu{\textbf{BLEU-3}} & \textbf{BLEU-4} & \textbf{ROUGE-L} &  \textbf{Meteor} & \textbf{Number} & \textbf{Ratio}\\
    \midrule
    \textbf{1-10}  & \textbf{25.04} & \textbf{16.89} & \textbf{13.22} & \textbf{11.37 } & \textbf{24.44} & \textbf{20.35} & 1996 & 13.60\%  \\
    \textbf{10-30} & 22.62 &14.70 & 11.58 & 10.10 & 21.98 & 17.79 & 3435 & 23.41\%  \\
    \textbf{30-50} &23.16 &14.62 &10.96 & 9.17  & 22.86 & 18.50 & 2652 & 18.07\%   \\
    \textbf{50-100} &24.26 &16.02 &12.53 & 10.87  & 23.65 & 19.28 & \textbf{3670} & \textbf{25.01}\% \\
    \textbf{100+}  & 21.63 & 13.75 & 10.68 & 9.24 & 22.52 & 17.70 & 2921 & 19.90\%  \\
    
    \bottomrule
	\end{tabular}
    \end{threeparttable}
    }
\end{table*}

\subsection{Our Benchmark} \label{sec:benchmark}
We crawl our benchmark from 56 popular java projects ranked by ``star numbers``. We have devoted substantial efforts to clean the dataset and compared with Jiang's dataset~\cite{DBLP:conf/kbse/JiangAM17}, ours is able to serve as more research purposes.

Specifically, we store commits in a format file with various attributes including ``commit\_id``, ``subject``, ``commit message``, ``\texttt{diff}``, and ``file\_changed``. Note that the ``subject`` refers to the first sentence extracted from the commit messages, which can be seen as the summary of a message\cite{DBLP:conf/kbse/JiangAM17}. ``file\_changed`` is the number of files that the current commit made.

Moreover, we also provide the extracted \textit{added}/\textit{deleted} functions from commits. For each commit, we extract the related functions. We name these functions in a format of ``project\_id\_positive(negative)\_num.java``, where ``project`` represents the commit belonging to which project, ``id`` is the hash value and ``positive/negative`` denotes the \textit{added}/\textit{deleted} functions and ``num`` is the number of extracted functions.

Finally, the noisy commits\cite{DBLP:conf/kbse/LiuXHLXW18}, e.g., bot messages, which refers to messages generated by development tools and trivial messages, which contains little information, have been filtered automatically, since we keep commits modified in \textit{.java} files and these boot messages and trivial messages most exist in configuration files, e.g., \textit{*.md}, \textit{*.gitrepo}. Furthermore, we remove the same content of the commits to ensure the benchmark \liu{has} a higher quality compared with Jiang's~\cite{DBLP:conf/kbse/JiangAM17}.

The benchmark contains the basic commit information and the completed functions altered by commits. Hence with this dataset, we can boost some other researches, e.g., code summarization, code recommendation, knowledge graph construction based on commits. 
\begin{tcolorbox}[breakable,width=\linewidth,boxrule=0pt,top=1pt, bottom=1pt, left=1pt,right=1pt, colback=gray!20,colframe=gray!20]
\textbf{} The benchmark we prepared contains adequate information as compared to Jiang's ~\cite{DBLP:conf/kbse/JiangAM17}, and we make it publicly \sq{available} \cite{Commit_data} to benefit community research.
\end{tcolorbox}

\subsection{Novelty of \tool}
\sql{Our generation module \textit{AST2seq} is inspired by Code2seq~\cite{alon2018code2seq}, however, it has the following major differences.}

\begin{itemize}
    \item \sql{\textbf{Input Handling.} Although Code2seq adopts ASTs to encode source code for tasks such as code captioning, code documentation, and code summarization, it exploits function-level ASTs. For the proposed \textit{AST2seq}, only partial function fragments, i.e., code changes, are considered for generating commit messages, which is more challenging. To construct ASTs for the code changes, we retrieve the whole function, including both \textit{added} and \textit{deleted} functions, and extract the AST paths corresponding to the changed code. To the best of our knowledge, we are the first to incorporate ASTs for the commit message generation.}
    
    %  Task Aspect. Code2seq used AST to encode the completed functions for some tasks e.g., code captioning, code documentation, code summarization. The granularity was at function level. However, our task targets at commit message generation and the data utilized are code changes, which are the partial code of the completed functions with \textit{added} and \textit{deleted} groups. Hence, to construct corresponding ASTs, we need to retrieved the completed functions and parse the AST paths for the changed code i.e., \textit{add/delete} code.  The complexity is higher than codeseq and to the best of our knowledge, we are the first to use AST in the commit message generation.
    \item \sql{\textbf{Model Design.} Code2seq employs a bi-directional LSTM to encode function-level ASTs. For \textit{AST2seq}, \textit{added} and \textit{deleted} code fragments are treated separately. \textit{AST2seq} first learns the representations of \textit{added} and \textit{deleted} code based on their respective ASTs by using two bi-directional LSTMs. Then the code change representations, i.e., \texttt{diff}s, are obtained by concatenating the two learned features.} 
    
    % Network Architecture. Code2seq employed a bi-directional LSTM to encode AST paths of completed functions, however, in our scenario, based on code changes i.e., \textit{added} code and \textit{deleted} code, we optimize to use two bi-directional LSTMs with different learnable parameters to learn \textit{added} and \textit{deleted} AST paths separately. Finally, the learned features are concatenated to represent the entire code changes i.e., \texttt{diff}s.
\end{itemize}

\begin{tcolorbox}[breakable,width=\linewidth,boxrule=0pt,top=1pt, bottom=1pt, left=1pt,right=1pt, colback=gray!20,colframe=gray!20]
\textbf{} \sql{In summary, although both \textit{AST2seq} and Code2seq utilize ASTs and bi-directional LSTM to learn code representations, they are different in input code handling and model design.}
\end{tcolorbox}

\subsection{Influence of Parameter Setting}
\begin{figure}[t!]
    \centering
    \begin{subfigure}[t]{0.22\textwidth}
        \centering
        \includegraphics[height=1.2in]{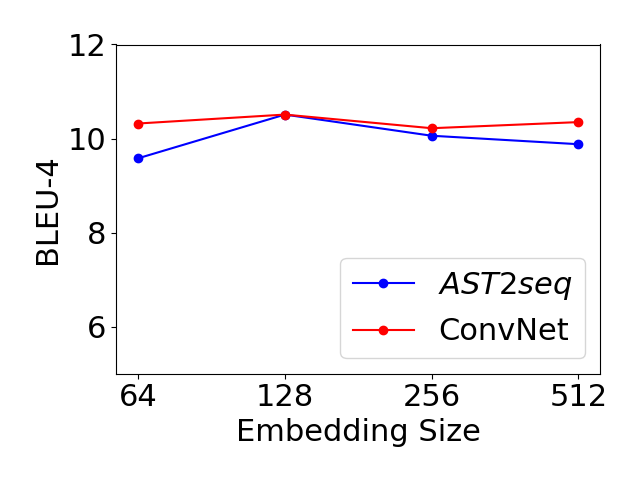}
        \caption{\sql{Embedding size of \textit{Ast2seq} and ConvNet.}}
        \label{fig:embedding_size}
    \end{subfigure}%
    ~ 
    \begin{subfigure}[t]{0.22\textwidth}
        \centering
        \includegraphics[height=1.2in]{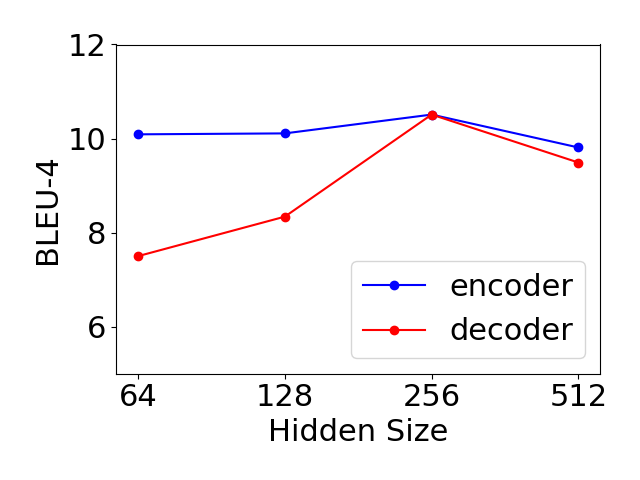}
        \caption{\sql{Hidden Size in encoder and decoder.}}
         \label{fig:encoder}
    \end{subfigure}
    \caption{\sql{Effect of different parameter settings.}}
    \label{fig:hyper-parameter}
\end{figure}

\sq{Besides the number of AST paths (discussed in Section 4.3.3), we also analyzed the influence of other parameters, including the the embedding size~\cite{siow2020core} in \textit{AST2seq} and ConvNet, and the number of hidden size~\cite{siow2020core} in encoder and decoder, on the model performance. The experimental results are depicted in Fig.~\ref{fig:hyper-parameter}. As shown in Fig.~\ref{fig:hyper-parameter} (a), when the embedding sizes of \textit{AST2seq} and ConvNet equal to 128, \tool achieves the best performance. According to Fig.~\ref{fig:hyper-parameter} (b), the number of hidden size in decoder can influence the model performance more obviously than the parameter in encoder. For example, \tool shows a dramatic increase when the number of hidden size in the decoder increases from 64 to 256; however, the corresponding fluctuation for the parameter in encoder is marginal. We define the numbers of hidden size in encoder and decoder as 256 due to the good performance in the experiment.}

\subsection{OOV Issue}
\sq{Some existing studies~\cite{Ptr-Net_code, DBLP:conf/ijcai/Xu00GT019} utilize strategies such as copy mechanism~\cite{see2017get} for alleviating the Out-Of-Vocabulary (OOV) issue, i.e., some tokens in the test set have not appeared in the fixed vocabulary built during training. The copy mechanism~\cite{see2017get} can learn to copy some tokens directly from the input \texttt{diff}s instead of only generating tokens based on the fixed vocabulary, and thereby mitigating the OOV issue. However, for our benchmark dataset, the vocabulary built during training contains 90,969 unique tokens from the \texttt{diff}s and commit messages, and only 1,930/14,674 (13.2\%) samples in the test set involve the tokens not appearing in the vocabulary, i.e., the OOV tokens. 
% Furthermore, only 168 samples from these 1,930 samples contain out-of-vocabulary tokens, which show up in both messages and \texttt{diff}s. 
Based on the observation, we suppose that using strategies such as copy mechanism might not improve much on ATOM.}

\subsection{Limitations}
\subsubsection{Model Complexity}
\tool encodes code changes based on AST to represent code semantics and further designs a ranking module for more accurate commit message generation. It contains two modules involved with deep learning approaches, which cost time and efforts to tune the best models. The complexity of extracting AST paths from functions based on \texttt{diff}s is far more than treating \texttt{diff}s as sequences during the preprocessing. Furthermore, the output produced by the retrieval module is incorporated into the generation module to make the final decision, which is a complicated pipeline and the workload is much bigger than the previous work. Once \tool is applied to the new benchmark, we still need to spend time and efforts to finish the preprocessing and model tuning. This can 
be considered as a limitation of \tool, however, it is inevitable for all deep learning approaches and once \textit{AST2seq} and ConvNet are fixed with the best parameters, the generation process is relatively low-cost and convenient. 

\sql{We provide a deep analysis on the efficiency by comparing \tool with the best retrieval-based approach, NNGen, and the best generation-based approach, NMT\textsubscript{(Luong)}. The comparison experiments are conducted on the same server with 36 cores and Nvidia Graphics Tesla P40 with 22GB memory. The comparison results are listed in Table~\ref{tbl-time}. As can be seen from the table, \tool costs more time on training and testing than NMT\textsubscript{(Luong)}, which reflects the more complexity of \tool than NMT\textsubscript{(Luong)}. Although training is unnecessary for NNGen, it spends the most time (308 secs) on testing since GPU cannot be used for acceleration.}

\begin{tcolorbox}[breakable,width=\linewidth,boxrule=0pt,top=1pt, bottom=1pt, left=1pt,right=1pt, colback=gray!20,colframe=gray!20]
\textbf{} \sql{Although \tool takes much time to tune the hyperparameters to get a best model, once the model is fixed, its application is efficient.}
\end{tcolorbox}

\begin{table}[t]
  \centering
%   \addtolength{\tabcolsep}{-3.5pt}
	\caption{\sql{Time costs of NNGen, NMT\textsubscript{(Luong)} and \tool. Since NNGen does not need training, its training time is marked as ``N/A''.}}
	\label{tbl-time}
	\scalebox{0.87}{\begin{threeparttable}
	\begin{tabular}{c | c c c }
	  \toprule 
    \textbf{Methods} & \textbf{Device} & \textbf{Training Time} &
     \textbf{Testing Time} \\
    \midrule
     NNGen & CPU & N/A &  308 secs\\
     NMT\textsubscript{(Luong)}  & Tesla P40 & 11 hours & 188 secs\\
     ATOM  & Tesla P40 & 16 hours & 257 secs  \\
    \bottomrule
	\end{tabular}
    \end{threeparttable}
    }
\end{table}

\subsubsection{Dataset Partition}
\textbf{Split by project.} 
In this paper, we follow the prior studies on commit message generation~\cite{liu2019generating, DBLP:conf/ijcai/Xu00GT019, DBLP:conf/kbse/JiangAM17,DBLP:conf/acl/LoyolaMM17, jiang2019boosting} by splitting dataset by commit. 
According to LeClair and McMillan's study on code summarization~\cite{leclair2019recommendations}, splitting dataset by ``function'' (in analogy with ``commit'' in our study) might cause information leakage from test set projects into the training or validation sets and should be avoided. 
Following the study, we evaluate the performance of \tool based on the dataset split by project instead of by commit. 
% \liu{Furthermore, since commits have the attribute of ``timestamp'', we also conduct an experiment to evaluate the performance of \tool split by timestamp.}
We compare \tool with the best retrieval-based model NNGen and the best generation-based model NMT\textsubscript{(Luong)} in the study.

The comparison results are illustrated in Table~\ref{tbl-project-1}. We find that \tool outperforms baseline models and the performance of all models decreases \liu{compared} to the performance when splitting dataset by commit. Although a reduced performance is reasonable and expectable based on the dataset split by project~\cite{leclair2019recommendations}, the magnitude of the decline is extremely obvious in our scenario. This indicates that splitting dataset by project may not be applicable for the evaluation of the commit message generation task. We further analyze the reason behind the extremely poor performance when splitting dataset by project from three aspects: method, task, and benchmark.

\begin{itemize}
    \item \sql{\textbf{Method.} Deep learning-based models generally require massive data to learn the prior knowledge. When splitting the dataset by project, no prior knowledge of the project in the testset will be learned during training. Thus, the performance are expected to be worse than dataset partition by commit.}
    % Model Aspect. The deep learning-based approaches require massive data to learn the prior knowledge. However, if the dataset is split by ``project'', when the trained model is used on the unknown project, it has no prior knowledge to produce a promising performance. Hence, the decrease is reasonable.
    \item \sql{\textbf{Task.} The code summarization task studied in~\cite{leclair2019recommendations} is different from code commit message generation task. In code summarization, the Java projects in the experimental dataset adopt some similar functions, e.g., ``\textit{setter}'' and ``\textit{getter}''~\cite{jiang2019machine}. So this part of knowledge from other projects can be helpful for summarizing code of an unknown project. However, in code commit message generation, code changes in one project may not appear in other projects, which hinders the knowledge adaption from the other projects.}
    % Task Aspect. The Java projects used for source code summarzation tend to have some similar or even identical functions e.g., ``setter'', ``getter'' methods~\cite{jiang2019machine}, hence though splitting the dataset based on ``project'', the similar data are still remained between train and test to assure the generation performance. Furthermore, Code2seq~\cite{alon2018code2seq} evaluated the proposed approach with different tasks: 1) code summarization; 2) code captioning; 3) code documentation. The first task is target at predicting function name and split the data by ``project''. However, source code summarization aims to generate a natural language sentence to describe the functionality of the program and this task just predicted the function name, which cannot be considered as code summarization. The remaining tasks focus on translate code snippets into natural languages and we classify into code summarization. These works all split data by ``sample'' rather than ``project''. 
    \item \sql{\textbf{Benchmark.} In the benchmark dataset, we filter out bot messages, trivial messages, and the same samples (see Section 5.2). The filtered messages tend to possess similar templates, e.g., ``This commit renames a file'' and ``Changes to a package''. Removing these messages increase the difficulty of generating commit messages based on the dataset split by project.}
    % Benchmark Aspect. We filter out the bot messages, trivial messages and the same samples in the entire dataset to construct a more challenging benchmark for commit message generation (See Section 7.2). Since the bot messages tend to have common templates i.e., ``This commit renames some files'', ``Changes to package XXX'', by removing these samples to ensure a higher quality dataset. Hence, on our challenging benchmark, without the highly similar samples, if we divide dataset by ``project'', \tool fails to learn any useful information for generation. 
\end{itemize}

\liu{\textbf{Split by timestamp.} The dataset splitting strategy based on either commit (following the prior studies~\cite{jiang2017towards, liu2019generating, DBLP:conf/ijcai/Xu00GT019} or project would render the training and test sets mingle with commit messages written at different timestamps. That is, the commits in the training set may be written after some commits in the test set, causing the model to learn ``from the future''. Such scenario may be unrealistic since ``future data'' are unavailable in practice. To mitigate the issue, we adopt another dataset splitting strategy, i.e., according to the committed timestamps. Specifically, for each project, we rank the commits in chronological order, and treat the earliest 90\% as training set and the rest as test set.
% Since by our preprocess (Section 4.1.1 Experimental Benchmark), the dataset has no identical sample, the dataset constructed avoid ``learn from feature'' strictly.
Comparison results based on the new dataset splitting strategy is illustrated in Table~\ref{tbl-timestammp}. As can be seen, \tool outperforms the baseline approaches with respect to all the evaluation metrics. Besides, the achieved scores are relatively lower than when splitting the dataset by commit. The reduced performance may be because 
developers of one project generally write commits for code changes related to different functionalities at different timestamps.
% i.e., functionalities at different timestamps, 
For example, during a period of time, the developers may focus on enhancing one functionality of the project, so more commits related to this functionality are written; while for a later period, more commits for a different functionality are posted. Some examples are illustrated in this link~\cite{ATOM_examples}. Using past commits for training may hinder the trained model to produce an accurate result for a later commit since the later commit may be related to a new functionality. Moreover, the developers serving for one functionality may be changed to writing another different functionality. The changing commit styles may also influence the prediction accuracy of the trained model.
% in the first month,  these commits may serve as 
% ``complete the functionality A" and in the next month, it may become ``complete the functionality B". However, A and B have no overlap to make model harder produce an accurate result based on split by timestamp. Furthermore, the developers may be changed at different timestamps to make code style changed. All of these undeline reasons make model harder to produce promising results. 
% and it would be hard for the model trained on historical commits to produce an accurate result of a new commit. 
However, the data partition strategy can laterally verify the generalizability of a proposed model for code commit message generation task, and we encourage the future research to consider such data partition strategy during evaluation.}  

\begin{tcolorbox}[breakable,width=\linewidth,boxrule=0pt,top=1pt, bottom=1pt, left=1pt,right=1pt, colback=gray!20,colframe=gray!20]
\textbf{} \sql{To sum up, \tool shows superior effectiveness than baseline models when splitting dataset by project or timestamp. However, \sq{due to} the characteristics of the commits, the best practise to split dataset is based on commit.} 
%Since few similar code changes appear across projects, training on different projects would not be helpful for generating commit messages of unknown projects.}
\end{tcolorbox}

% \begin{table}[t]
%   \centering
% %   \addtolength{\tabcolsep}{-3.5pt}
% 	\caption{\sql{The performance of different approaches based on the dataset split by project.}}
% 	\label{tbl-project-1}
% 	\scalebox{0.87}{\begin{threeparttable}
% 	\begin{tabular}{c | c c c c c c }
% 	  \toprule 
%     \textbf{Methods} & \textbf{BLEU-4} & \textbf{ROUGE-1} &
%      \textbf{ROUGE-2} & \textbf{ROUGE-L} &  \textbf{Meteor}\\
%     \midrule
%      NNGen & 0.15 & 0.06 & 0.01 & 0.05 & 0.04 \\
%      NMT\textsubscript{(Luong)}  & 0.00 & 0.06 & 0.01 & 0.04 & 0.03 \\
%      ATOM  & 0.82  & 0.09  & 0.01 & 0.07 & 0.05 \\
%     \bottomrule
% 	\end{tabular}
%     \end{threeparttable}
%     }
% \end{table}

\begin{table}[t]
  \centering
%   \addtolength{\tabcolsep}{-3.5pt}
	\caption{\sql{The performance of different approaches based on the dataset split by project.}}
	\label{tbl-project-1}
	\scalebox{0.8}{\begin{threeparttable}
	\begin{tabular}{c | c c c c c c c c}
	  \toprule 
    \textbf{Methods} & \liu{\textbf{BLEU-1}} & \liu{\textbf{BLEU-2}} & \liu{\textbf{BLEU-3}} &
     \textbf{BLEU-4} & \textbf{ROUGE-L} &  \textbf{Meteor}\\
    \midrule
     NNGen & 5.02 & 1.39 & 0.42 & 0.15  & 0.05 & 0.04 \\
     NMT\textsubscript{(Luong)}  & 4.48 & 0.98 &0.00 & 0.00  & 0.04 & 0.03 \\
     ATOM  & 5.44 & 2.06 & 1.28 & 0.82  & 0.07 & 0.05 \\
    \bottomrule
	\end{tabular}
    \end{threeparttable}
    }
\end{table}

\begin{table}[t]
  \centering
%   \addtolength{\tabcolsep}{-3.5pt}
	\caption{\sql{The performance of different approaches split by timestamp.}}
	\label{tbl-timestammp}
	\scalebox{0.8}{\begin{threeparttable}
	\begin{tabular}{c | c c c c c c c c}
	  \toprule 
    \textbf{Methods} & \liu{\textbf{BLEU-1}} & \liu{\textbf{BLEU-2}} & \liu{\textbf{BLEU-3}} &
     \textbf{BLEU-4} & \textbf{ROUGE-L} &  \textbf{Meteor}\\
    \midrule
     NNGen & 10.28 & 4.83 & 3.15 & 2.49  & 10.34 & 8.41 \\
     NMT\textsubscript{(Luong)}  & 6.69 & 2.70 & 1.25 & 1.47  & 9.45 & 6.11 \\
     ATOM  & 10.87 & 5.99 & 4.19 & 3.53  & 11.14 & 9.53 \\
    \bottomrule
	\end{tabular}
    \end{threeparttable}
    }
\end{table}

\subsection{Threats to Validity}
One of the threats to validity is about the collected dataset. Our dataset contains more information than Jiang's~\cite{DBLP:conf/kbse/JiangAM17} with more volumes,
but more data is always beneficial to deep learning models. With the dataset we crawled so far, we have already achieved the best performance, which indicates the effectiveness of our proposed approach. Another threat to validity is about human evaluation in Section~\ref{sec:human}. We ask 6 participants to evaluate the quality of 100 randomly selected commit messages according to the criterion~\cite{DBLP:conf/kbse/LiuXHLXW18}. However, we cannot guarantee the judgements of participants are fully in line with the criterion. Ideally, the scores obtained from 6 participants are more reliable than those labeled by 3 participants, which is a common strategy adopted by prior work~\cite{DBLP:conf/kbse/LiuXHLXW18}. 
\sql{Furthermore, the reproduction of NNGen~\cite{DBLP:conf/kbse/LiuXHLXW18} may introduce bias to the experimental results. To alleviate the threat, we have tried our best to read the paper carefully and consulted the authors about the details to ensure our reproduction is correct.
}
Finally, we only compare \tool on our dataset with the baseline methods and get state-of-the-art performance. As Jiang's~\cite{DBLP:conf/kbse/JiangAM17} dataset does not provide commit ids, we cannot extract the \textit{Added} and \textit{Deleted} ASTs to encode changes. Hence, we cannot verify the effectiveness of \tool on Jiang's dataset. But we compared all existing generation and retrieval approaches with ours on our benchmark to illustrate the effectiveness of \tool.

\section{Related Work}\label{sec:related}
%\sen{Will start to edit the related work  tomorrow.}
{Our work is inspired by two research lines of studies, including code commit message generation and code summarization.
In this section, we discuss the most related work and compare them with \tool.}

\subsection{Code Commit Message Generation}
%\noindent{\bf Rule-based CCMG.} 
Previous commit message generation studies can be mainly categorized into three types according to the methodology: rule-based, retrieval-based, and deep-learning-based. 
Initial studies~\cite{DBLP:conf/kbse/BuseW10,DBLP:conf/scam/Cortes-CoyVAP14,DBLP:conf/icse/VasquezCAP15,DBLP:conf/compsac/ShenSLYH16} rely on pre-defined rules or templates to establish the connections between code changes and natural languages. 
For example, Buse et al.~\cite{DBLP:conf/kbse/BuseW10} use the templates based on control flows to generate commit messages. Shen et al.~\cite{DBLP:conf/compsac/ShenSLYH16} extract code changes based on defined types of changed methods and corresponding formats (e.g., ``replace \textless old method name\textgreater\,
with \textless new method name\textgreater'' is a defined format for renaming a method). ChangeSribe \cite{DBLP:conf/scam/Cortes-CoyVAP14,DBLP:conf/icse/VasquezCAP15} further takes the impact set of a commit into account along with the commit stereotype and type of changes using pre-defined metrics, then fills a pre-defined commit message template with the extracted information.
Such rule-based approaches can be limited by the manually specified rules or templates, and may work inefficiently for the code changes not applicable to the rules.

The retrieval-based approaches~\cite{DBLP:conf/esem/HuangZCXLL17,DBLP:conf/kbse/LiuXHLXW18} regard a newly-arrived \texttt{diff} as a query and reuse the commit messages of the most similar code changes. Huang et al. \cite{DBLP:conf/esem/HuangZCXLL17} use the syntactic similarity and semantic similarity of code changes as a measurement to retrieve existing commit messages. NNGen \cite{DBLP:conf/kbse/LiuXHLXW18} reuses the message of the nearest neighbour by computing the cosine similarity of \texttt{diff} vectors constructed by a bag-of-words model, which extends to include both codes changes and non-code changes.
For these approaches, simply retrieving messages as the targets cannot guarantee the consistency of the variable/method names. Besides, the mapping relations between \texttt{diff}s and commit messages are not fully exploited.

Deep-learning-based approaches~\cite{DBLP:conf/kbse/JiangAM17,DBLP:conf/acl/LoyolaMM17,DBLP:conf/ijcai/Xu00GT019} treat code changes and commit messages as two different languages, and design neural machine translation (NMT) models to translate code changes into commit messages. For example, Jiang et al.~\cite{DBLP:conf/kbse/JiangAM17} directly adopt NMT model to conduct the translation. 
\sql{Jiang et al.~\cite{jiang2019boosting} also adopt NMT model to conduct the translation but with all the commit messages formatted based on ChangeScribe~\cite{DBLP:conf/scam/Cortes-CoyVAP14,DBLP:conf/icse/VasquezCAP15}.}
CODISUM~\cite{DBLP:conf/ijcai/Xu00GT019} propose to combine both code structure and code semantics to enrich the representations of code changes for a better generation, and use CopyNet to mitigate the OOV issue. Although the results for these approaches are promising, they still do not explicitly bridge the gap between code and natural languages. 

Compared with the above works, \tool encodes ASTs to represent code changes and fully takes advantages from both retrieved methods and deep-learning-based methods by involving a hybrid ranking module to boost the performance further, resulting in more accurate commit message generation than all the above works.  

\subsection{Code Summarization}
Code summarization aims to generate brief natural language descriptions for code snippet and it evolves from rule-based~\cite{abid2015using}~\cite{sridhara2010towards}~\cite{moreno2013automatic}, retrieval-based~\cite{haiduc2010supporting}\cite{haiduc2010use}~\cite{wong2015clocom} to learning-based~\cite{allamanis2016convolutional}~\cite{iyer2016summarizing}~\cite{wan2018improving} approaches. Pre-defining some basic rules based on the important content from codes is one of the most common approaches for the generation. Sridhara et al.~\cite{sridhara2010towards} design a framework with traditional program and natural language analysis to tokenize function/variable names to summarize the Java method. Furthermore, based on this framework, Moreno et al.~\cite{moreno2013automatic} \liu{predefine} rules to combine information to generate comments for Java classes.

Information retrieval approaches are widely used in summary generation tasks. Haiduc et al.~\cite{haiduc2010supporting} use Vector Space Model (VSM) and Latent Semantic Indexing (LSI), an information retrieval method, to index top-k terms from a function and find the most similar terms based on cosine distances as the summary. Rodeghero et al.~\cite{Rodeghero} further improve the performance by improving the subset selection process by modifying the weights of the keywords from the codes based on the result of an eye-tracking study. McBurney et al.~\cite{mcburney2014improving} apply topic \liu{modelling} and design a hierarchy to organize the topics in source code, with more general topics near the top of the hierarchy to select keywords and topics as code summaries. Clocim~\cite{wong2015clocom} applies code clone detection to find similar codes and uses its comments directly.

In addition, some researchers try to generate summaries by learning-based approaches. Iyer et al.~\cite{iyer2016summarizing} propose CODE-NN, an attentional LSTM encoder-decoder network to generate C\# and SQL descriptions. Hu et al.~\cite{xinghu} further incorporate an additional encoder layer into the NMT model to learn API sequence knowledge. They first train an API sequence encoder using an external dataset, then apply the learned representation into the encoder-decoder model to assist generation.
Wan et al.~\cite{wan2018improving} also incorporate an abstract syntax tree as well as sequential content of code snippets into a deep reinforcement learning framework to translate python code snippets. Code2seq~\cite{alon2018code2seq} model represents a code snippet as the set of paths in its AST to decode language sequences and the results outperform state-of-the-art NMT models. Different from code summarization, we aim at generating code changes, a higher target compared with the whole function summary.

\section{Conclusion}\label{sec:conclusion}
Automatically generating commit messages is necessitated. Existing studies either translate \texttt{diff}s with sequence-based methods or retrieval-based methods. In this paper, we propose our \tool to encode AST paths of \texttt{diff}s for code representation to generate commit messages. Furthermore, we integrate the advantages of retrieval-based models by a hybrid ranking module to prioritize the most accurate message from both retrieved and generated messages. Substantial experiments based on our benchmark have demonstrated the effectiveness of \tool and \tool increases the state-of-the-art approaches by \textbf{30.72\%} in BLEU-4. In future work, we plan to design a detailed specification to keep commits with higher quality and apply our proposed approach to other tasks such as code summarization, code documentation, and even source code generation.

% \input{sections/ack.tex}
% \input{sections/bio.tex}
% \clearpage
\balance
\bibliographystyle{IEEEtran}
\bibliography{reference.bib}
\end{document}